\documentclass[reprint,superscriptaddress,amsmath,amssymb]{revtex4-2}

\usepackage{graphicx}% Include figure files
\usepackage{dcolumn}% Align table columns on decimal point
\usepackage{bm}% bold math
\usepackage[driverfallback=dvipdfmx,colorlinks,linkcolor=blue,citecolor=blue,urlcolor=blue]{hyperref}% add hypertext capabilities
\usepackage{ulem}
\usepackage[english]{babel}
\usepackage{svg}
\usepackage{changes}

\newcommand{\rw}[1]{\textcolor{black}{#1}}
\begin{document}

%\title{Quantum control and \replaced{Berry phase of electron spins in }{fast rotation of} levitated diamonds in high vacuum}
\title{Quantum control and Berry phase of electron spins in rotating levitated diamonds in high vacuum}

    \author{Yuanbin Jin}%
    \thanks{These authors contributed equally to this work.}
\affiliation{	Department of Physics and Astronomy, Purdue University, West Lafayette, Indiana 47907, USA}

	\author{Kunhong Shen}%
	 \thanks{These authors contributed equally to this work.}
\affiliation{	Department of Physics and Astronomy, Purdue University, West Lafayette, Indiana 47907, USA}

	\author{Peng Ju}
\affiliation{	Department of Physics and Astronomy, Purdue University, West Lafayette, Indiana 47907, USA}

	\author{Xingyu Gao}
\affiliation{	Department of Physics and Astronomy, Purdue University, West Lafayette, Indiana 47907, USA}

	\author{Chong Zu}
\affiliation{Department of Physics, Washington University, St. Louis, MO, 63130, USA}

	\author{Alejandro J. Grine}
\affiliation{Sandia National Laboratories, P.O. Box 5800, Albuquerque, NM, 87185, USA}

	\author{Tongcang Li}
\email{tcli@purdue.edu}
\affiliation{	Department of Physics and Astronomy, Purdue University, West Lafayette, Indiana 47907, USA}
\affiliation{	Elmore Family School of Electrical and Computer Engineering, Purdue University, West Lafayette, Indiana 47907, USA}
\affiliation{	Purdue Quantum Science and Engineering Institute, Purdue University, West Lafayette, Indiana 47907, USA}
\affiliation{	Birck Nanotechnology Center, Purdue University, West Lafayette, Indiana 47907, USA}

\date{\today}

\begin{abstract}
Levitated diamond particles in high vacuum with internal spin qubits have been proposed for exploring macroscopic quantum mechanics, quantum gravity, and precision measurements. The coupling between spins and particle rotation can be utilized to study quantum geometric phase, create gyroscopes and rotational matter-wave interferometers. However, previous efforts in levitated diamonds struggled with vacuum level or spin state readouts. To address these gaps, we fabricate an integrated surface ion trap with multiple stabilization electrodes. This facilitates on-chip levitation and, for the first time, optically detected magnetic resonance measurements of a nanodiamond levitated in high vacuum. The internal temperature of our levitated nanodiamond remains moderate below $10^{-5}$ Torr.  Impressively, we have driven a nanodiamond to rotate up to 20 MHz ($1.2 \times 10^{9}$ rpm), surpassing \rw{typical} nitrogen-vacancy (NV) center electron spin dephasing rates. Using these NV spins, we observe the \rw{effect of the Berry phase} arising from particle rotation. In addition, we demonstrate quantum control of spins in a rotating nanodiamond. These results mark an important development in interfacing mechanical rotation with spin qubits, expanding our capacity to study quantum phenomena.

\end{abstract}

\maketitle

\section{Introduction}

Levitated nanoparticles and microparticles in high vacuum \cite{gonzalez2021levitodynamics,millen2020optomechanics,winstone2023levitated} offer a remarkable degree of isolation from environmental noises, rendering them exceptionally suitable for studying fundamental physics \cite{Romero-Isart2011,Afek2022,Yin2022}  and executing precision measurements \cite{Geraci2010,Hebestreit2018,Hoang2016PRL,zheng2020robust,zhu2023nanoscale}. Recently, the center-of-mass (CoM) motion of levitated nanoparticles in high vacuum has been cooled to the quantum regime \cite{Delic2020,Magrini2021,Tebbenjohanns2021}. Unlike tethered oscillators, levitated particles can also exhibit rotation \cite{Arita2013,Kuhn2017,Reimann2018,Ahn2018,Ahn2020,Jin2021,zeng2023optically,ju2023near}, which is intrinsically nonlinear \cite{ma2020quantum,stickler2021quantum}. Beyond rigid-body motion, levitated particles can \rw{host} embedded spin qubits to provide more \rw{functionalities } \cite{Yin2013,Scala2013}. Notably, levitated nanodiamonds with embedded nitrogen-vacancy (NV) center spin qubits have been proposed for creating massive quantum superpositions \cite{Yin2013,Scala2013} to test the limit of quantum mechanics and quantum gravity \cite{bose2017spin,marletto2017gravitationally}. The embedded spin qubits can also sense the pseudo-magnetic field \cite{Wood2017,Wood2018,Chudo2014}, \rw{related to} the Barnett effect \cite{Barnett1915,Barnett1935}, and the quantum geometric phase \cite{Maclaurin2012,Chen2019} \rw{associated with} particle rotation. The coupling between spin and mechanical rotation can be utilized for building sensitive gyroscopes \cite{Ledbetter2012,zhang2023highly} and rotational matter-wave interferometers \cite{ma2017proposal,Rusconi2022}. These innovative proposals necessitate levitating diamond particles in a high vacuum, well below 10$^{-3}$ Torr. However, prior experiments with levitated diamonds struggled with vacuum level or spin state readouts. 

\begin{figure*}[htp]
	\includegraphics[width=0.98\textwidth]{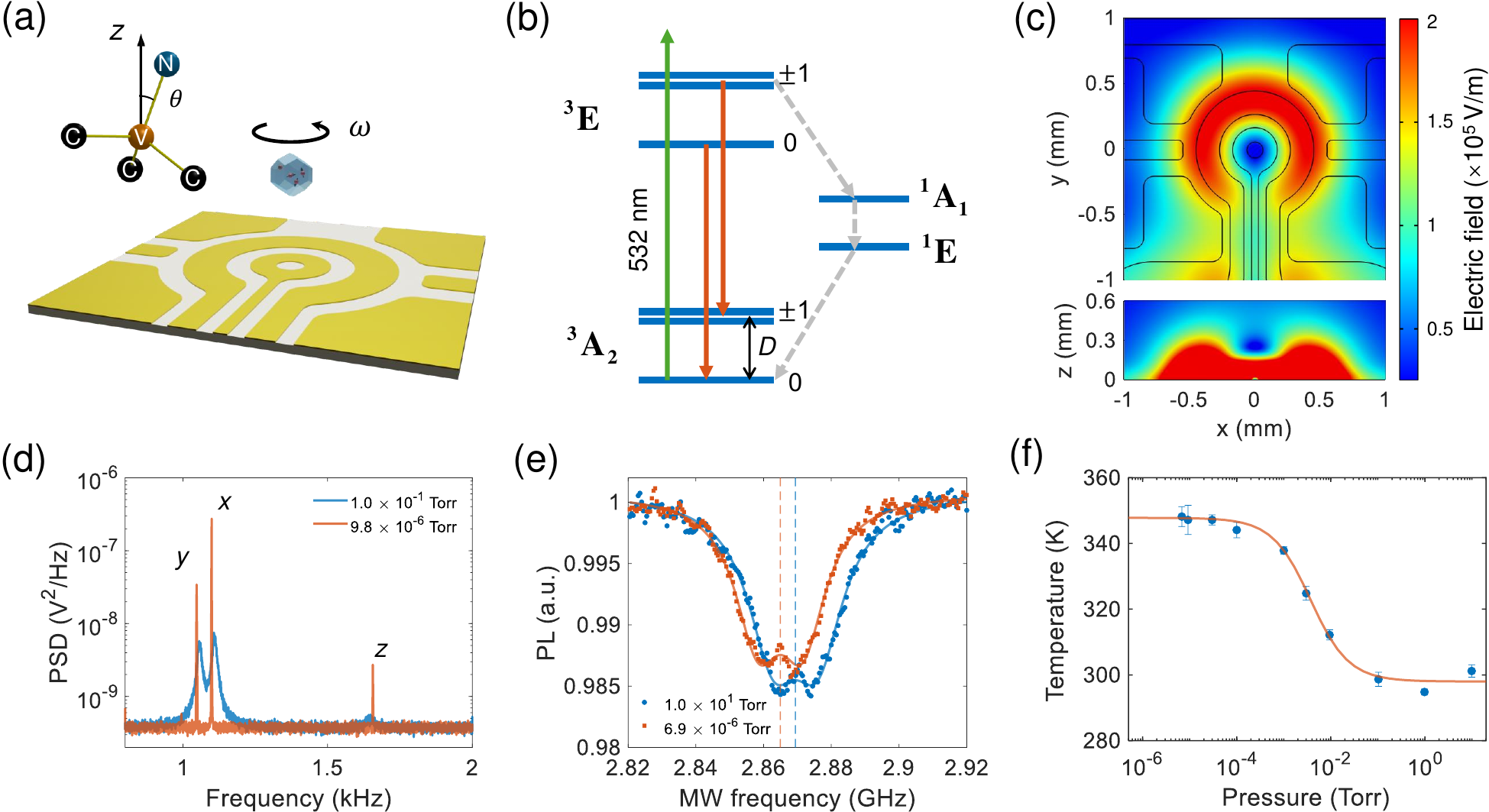}
	\caption{\label{fig:1} Stable levitation of a nanodiamond in high vacuum. (a) Schematic of a levitated nanodiamond in a surface ion trap. The center ring electrode is grounded (GND). It has a hole at its center for sending a 1064 nm laser to monitor the nanodiamond's motion. A combination of a low-frequency high voltage (HV) and a high-frequency microwave (MW) is applied to the $\Omega$-shaped circuit to trap the nanodiamond and control the NV centers. (b) Energy level diagram of a diamond NV center. A 532 nm laser (green arrow) excites the NV center. The red solid arrows and gray dashed arrows represent radiative decays and nonradiative decays, respectively. (c) Simulation of the electric field of the ion trap in the $xy$-plane (top) and in the $xz$-plane (bottom) when a voltage of 200 V is applied to the $\Omega$-shaped circuit. The trap center is 253 $\mu$m away from the chip surface. (d) Power spectrum densities (PSDs) of the center-of-mass (CoM) motion of the levitated nanodiamond at the pressure of 0.1 Torr (blue) and $9.8 \times 10^{-6}$ Torr (red). (e) Optically detected magnetic resonances (ODMRs) of the levitated nanodiamond measured at 10 Torr (blue circles) and $6.9 \times 10^{-6}$ Torr (red squares). The blue and red dashed lines are the corresponding zero-field splittings. The intensities of the 532 nm laser and the 1064 nm laser are 0.030 W/mm$^2$ and 0.520 W/mm$^2$, respectively. (f) Internal temperature of the levitated nanodiamond as a function of pressure with the same laser intensities as shown in (e).}
\end{figure*}

Optical levitation of nanodiamonds has been experimentally achieved, but it was restricted to pressures above 1~Torr due to laser-induced heating \cite{neukirch2015multi,Hoang2016,Frangeskou_2018}. Earlier studies using Paul traps, or ion traps, have demonstrated spin cooling \cite{Delord2020} and angle locking \cite{perdriat2022angle} of levitated diamonds. However, they encountered a similar issue: diamond particles were lost when the air pressure was \rw{reduced} to about 0.01~Torr \cite{delord2017diamonds,conangla2018motion,Delord2020,perdriat2022angle,perdriat2023spin}. This phenomenon could be  due to the nonideal design of the Paul traps used in those experiments or the heating effects of detection lasers. While nanodiamonds can be levitated in a magneto-gravitational trap \cite{hsu2016cooling,OBrien2019}, reading out the spin state within this setup remains elusive. 

In this article, we design and fabricate an integrated surface ion trap (Fig. \ref{fig:1}(a)) that incorporates an $\Omega$-shaped \rw{stripline to deliver} both a low-frequency high voltage  for trapping and a microwave  for NV spin control. Additionally, it \rw{comprises} multiple electrodes to stabilize the trap and drive a levitated diamond to rotate. With this advanced Paul trap, we have performed optically detected magnetic resonance (ODMR) measurements of a levitated nanodiamond in high vacuum for the first time. Using NV spins, we measure the internal temperature of the levitated nanodiamond, which remains stable at approximately 350~K under  pressures below $10^{-5}$~Torr. This suggests prospects for levitation in ultra-high vacuum. With a rotating electric field, we have been able to drive a levitated nanodiamond to rotate at high speeds up to 20 MHz ($1.2 \times 10^{9}$ rpm), which is about three orders of magnitudes faster than \rw{previous} achievements using diamonds mounted on  motor spindles \cite{Wood2017,Wood2018}. Notably, this rotation speed surpasses the \rw{typical} dephasing rate of NV spins in the diamond. With embedded NV electron spins in the levitated nanodiamond, we observe the \rw{effect of Berry phase} generated by the mechanical rotation, which also improves the ODMR spectrum in an external magnetic field. Moreover, we achieve quantum control of NV centers in a rotating levitated nanodiamond. Our work represents a pivotal advancement in interfacing mechanical rotation with spin qubits.

\section{Results}

\subsection{Levitation of a nanodiamond in high vacuum}

In the experiment, we levitate a nanodiamond in vacuum using a surface ion trap (Fig. \ref{fig:1}(a)). The surface ion trap is fabricated on a sapphire wafer, \rw{which has} high transmittance for visible and near-infrared lasers. To achieve levitation of nanodiamonds and quantum control of NV spins simultaneously, we apply both an AC high voltage and a microwave on a $\Omega$-shaped circuit. The AC high voltage has a frequency of about 20 kHz and an amplitude of about 200 V. The microwave has a frequency of a few GHz.  They are combined together with a homemade bias tee. The center ring electrode is grounded to generate a trapping center above the chip surface. The four electrodes at the corners are used to compensate the static electric fields from surface charges to minimize the micro-motion of a levitated nanodiamond. Fig. \ref{fig:1}(c) shows a simulated distribution of the electric field of the trap. The trapping center is 253 $\mu$m away from the chip surface. 

The trapping potential depends on the charge to mass ratio ($Q/m$) of a levitated particle. Thus, it is necessary to increase the charge number of particles for stable levitation in an ion trap. In our experiment, nanodiamonds are charged and sprayed out by electrospray and delivered to the surface ion trap with an extra linear Paul trap. The charge of the sprayed nanodiamond is typically larger than 1000 $e$, where $e$ is the elementary charge, enabling a large trapping depth of more than 100 eV (see Supplementary Information for more details). \rw{A 532 nm laser is used} to excite diamond nitrogen-vacancy (NV) centers and a 1064 nm laser \rw{is applied} to monitor the nanodiamond's motion. More details of our experimental setup are shown in Supplementary Fig. 1.

A main result of our experiment is that we can levitate a nanodiamond with the surface ion trap in high vacuum, which is a breakthrough as levitated diamond particles were lost around 0.01~Torr in previous studies using ion traps \cite{delord2017diamonds,conangla2018motion,Delord2020,perdriat2022angle,perdriat2023spin}. The red curve in Fig. \ref{fig:1}(d) shows the power spectrum density (PSD) of the center-of-mass (CoM) motion of a levitated nanodiamond at $9.8 \times 10^{-6}$ Torr. The radius of the levitated nanodiamond is estimated to be about 264 nm based on its PSDs at 0.01 Torr (Supplementary Fig. 2).  Our surface ion trap is remarkably stable in high vacuum. We can levitate a nanodiamond in high vacuum continuously for several weeks.

The internal temperature of a levitated nanodiamond is important as it will affect the spin coherence time and trapping stability. We measure the internal temperature using NV centers. The energy levels of an NV center is shown in Fig. \ref{fig:1}(b). We use a 532 nm laser to excite the NV centers and a single photon counting module to detect their photoluminescence. Then we sweep the frequency of a microwave to perform the ODMR measurement of a levitated nanodiamond in the absence of an external magnetic field. Fig. \ref{fig:1}(e) shows the ODMRs measured at 10 Torr (blue circles) and $6.9 \times 10^{-6}$ Torr (red squares). Based on the fitting of the ODMRs, the corresponding zero-field splittings (blue and red dashed lines) are 2.8694 GHz and 2.8650 GHz, respectively. \rw{The} internal temperature of the levitated nanodiamond \rw{can be obtained form the zero-field splitting} (see Methods for details). The measured internal temperature at different pressures are shown in Fig. \ref{fig:1}(f). The internal temperature is close to the room temperature at pressures above 0.1 Torr, and increases when we reduce the pressure from 0.1 Torr to $10^{-4}$ Torr. Finally, it remains stable at approximately 350 K at pressures below $5 \times 10^{-5}$ Torr.  This temperature is low enough to maintain quantum coherence of NV spins for quantum control \cite{Toyli2012PhysRevX}. 

The observed phenomena (Fig. \ref{fig:1}(f)) arise from the balance between laser-induced heating (Supplementary Fig. 3) and the cooling effects of air molecules and black-body radiation on the internal temperature of a levitated nanodiamond \cite{Liu2006,Chang2010}. When the air pressure is high, the cooling rate due to surrounding air molecules is large and the internal temperature of the levitated nanodiamond is close to the room temperature. However, as air pressure decreases, cooling from air molecules diminishes, leading to a rise in internal temperature. When the pressure is below $5 \times 10^{-5}$ Torr, the temperature stabilizes as the cooling is dominated by the black body radiation, which is independent of the air pressure.

\subsection{Fast rotation and \rw{Berry phase}}

\begin{figure*}[ht]
	\includegraphics[width=0.98\textwidth]{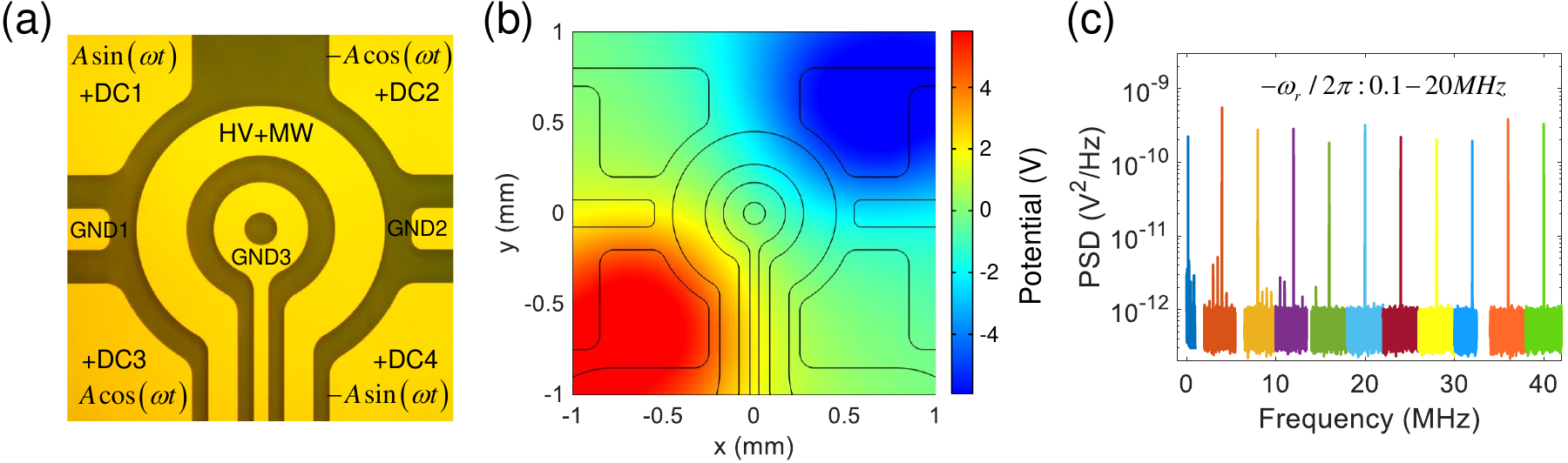}
	\caption{\label{fig:2} Fast rotation of a levitated nanodiamond. (a) Optical image of the surface ion trap. AC voltage signals ($A \sin{(\omega t+\varphi)}$) with the same frequency ($\omega$) and amplitude ($A$) but different phases ($\varphi$) are applied to the four corner electrodes to generate a rotating electric field. The phase is different by $\pi/2$ between neighboring electrodes. DC1, DC2, DC3 and DC4 are compensation voltages that minimize the micromotion to stabilize the trap. (b) Simulation of the electric potential in the $z=253\ \mu$m plane at $t = 0$. The amplitude is $A = 10$ V. (c) PSDs of the rotational motion of the levitated nanodiamond at the rotation frequencies from 0.1 MHz to 20 MHz. The pressure is $1.0 \times 10^{-4}$ Torr.}
\end{figure*}

\rw{After a nanodiamond is levitated in high vacuum}, we use a rotating electric field to drive the levitated nanodiamond to rotate at high speeds, which also \rw{stabilizes} the orientation of the levitated nanodiamond. The four electrodes at the corners are applied with AC voltage signals ($A \sin{(\omega t+\varphi)}$) with the same frequency ($\omega$) and amplitude ($A$) but different phases ($\varphi$) to generate a rotating electric field (Fig. \ref{fig:2}(a)). The phases of neighboring signals are different by $\pi/2$. Fig. \ref{fig:2}(b) shows the simulation of the electric potential in the $xy$-plane at $t = 0$. More information can be found in Supplementary Information (Supplementary Fig. 4 and Supplementary Fig. 5). A levitated charged object naturally has an electric dipole moment due to inhomogeneous distribution of charges. \rw{In a rotating electric field, the levitated charged particle will rotate due to the torque produced by the interaction between the rotating electric field and the electric dipole of the particle}. Fig. \ref{fig:2}(c) depicts the PSDs of the rotation at different driving frequencies (0.1 MHz – 20 MHz). The maximum rotation frequency is 20 MHz in the experiment, which is limited by our phase shifters used to generate phase delays between signals on the four electrodes. This is about 3 orders of magnitudes faster than previous achievements using diamonds mounted on electric motor spindles \cite{Wood2017,Wood2018}. When the rotation frequency is 100 kHz, the linewidth of the PSD of the rotational signal is fitted to be about $9.9 \times 10^{-5}$ Hz (Supplementary Fig. 5(d)), which is limited by the measurement time. This shows \rw{that} the rotation is extremely stable and is locked to the driving \rw{electric} signal. With easy control and ultra-stability, this driving scheme enables us to adjust and lock the rotation of the levitated nanodiamond over a large range of frequencies (see Supplementary Information for more details).

\rw{The fast rotating diamond with embedded NV spins allows us to observe the effects of the Berry phase due to mechanical rotation. The Berry phase, also known as the geometric phase,  is a fundamental aspect of quantum mechanics with applications in multiple fields, including the topological phase of matter and the quantum hall effect \cite{Tycko1987,Zhang2005,Leek2007,Xiao2010,Chudo2021}. The Berry phase in the laboratory frame is equivalent to the pseudo-magnetic field (called the  Barnett field in \cite{Chudo2021}) in the rotating frame: $B_{\omega}=\omega_r/\gamma$, where $\gamma$ is the spin gyromagnetic ratio. In this work, the microwave source is fixed in the laboratory frame. Only the levitated diamond is rotating. So, we can observe the effect of the Berry phase due to rotation \cite{Chudo2021}. }

\begin{figure*}[htp]
	\includegraphics[width=0.98\textwidth]{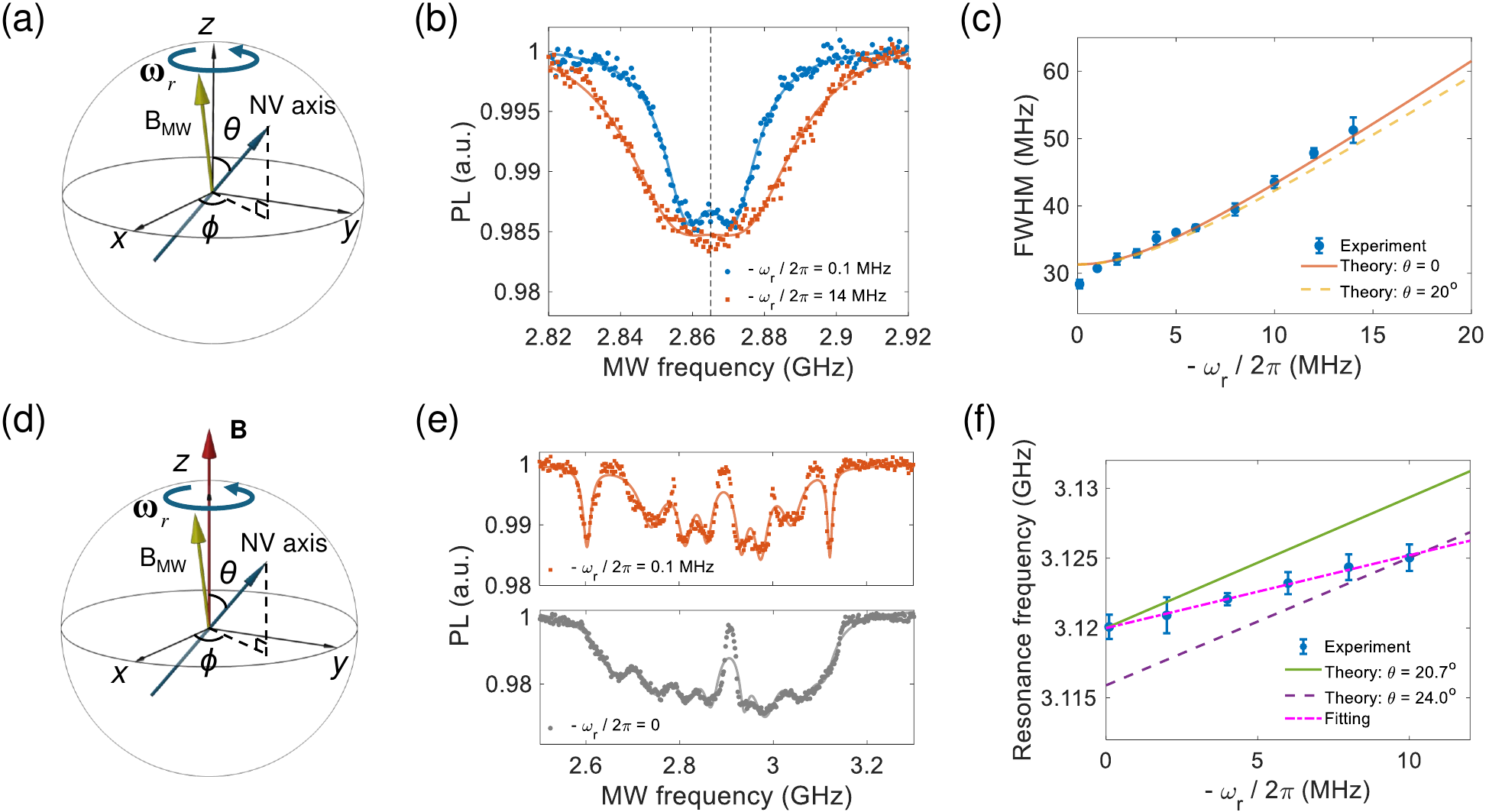}
	\caption{\label{fig:3} \rw{Effects of the Berry phase} generated by a rotating nanodiamond. \rw{(a) Schematic of an NV center in the nanodiamond rotating around the $z$ axis in the absence of an external magnetic field. The small angle between $B_{\mathrm{MW}}$ and the $z$ axis is due to the asymmetric design of the waveguide. (b) ODMRs of the levitated nanodiamond at rotation frequencies of 0.1 MHz (blue circles) and 14 MHz (red squares). (c) Experimentally measured FWHM of the ODMR spectrum as a function of rotation frequency (blue circles). The red solid curve and orange dashed curve are theoretically calculated FWHMs at $\theta = 0^ \circ$ and $\theta = 20^ \circ$, respectively.} (d) Schematic of an NV center in the nanodiamond rotating \rw{around} the $z$ axis in an external magnetic field. The magnetic field is along the $z$ axis and is about 100 G. (e) The upper panel (red squares) shows the ODMR of the levitated nanodiamond at a rotation frequency of 0.1 MHz and a pressure of $1.0 \times 10^{-4}$ Torr. The bottom panel (gray circles) shows the ODMR of a nanodiamond without a stable rotation at the pressure of 10 Torr. The corresponding solid curves are the fittings with eight Lorentzian dips. (f) Experimentally measured frequency of the right-most dip of the ODMR spectrum of a NV center as a function of rotation frequency (blue circles). The \rw{green} solid curve and \rw{violet} dashed curve are theoretical calculations at $\theta = 20.7^ \circ$ and $\theta = 24.0^ \circ$, respectively. \rw{The magenta dashed curve is a linear fitting of the resonance frequency.}}
\end{figure*}

\rw{In a rotating diamond, the embedded NV centers also follow the rotation  with an angular frequency of $\omega_r$ (Fig. \ref{fig:3}).} The levitated nanodiamond in our experiment contains ensembles of NV centers with four groups of orientations. \rw{Fig. \ref{fig:3}(d)} shows a NV center \rw{embedded} in a nanodiamond rotating \rw{around the} $z$ axis \rw{in the presence of an external magnetic field. The direction of the magnetic field is along the $z$ axis}. The angle between the NV axis and $z$ axis is $\theta$, and the azimuth is $\phi(t)$ relative to the \rw{$x$} axis. The Hamiltonian of the rotating NV electron spin \rw{in the laboratory frame, neglecting strain effects,} can be written as \cite{Maclaurin2012}:

\begin{widetext}
\begin{equation}
	\begin{array}{l}
		{H_{lab}} = {H_{0,lab}} + g{\mu _B}B{S_z} = \frac{1}{\hbar }R\left( t \right)DS_z^2{R^\dag }\left( t \right) + g{\mu _B}B{S_z}\\
		= D\hbar \left( {\begin{array}{*{20}{c}}
				{{{\cos }^2}\theta  + \frac{{{{\sin }^2}\theta }}{2} + \frac{{g{\mu _B}B}}{D}}&{\frac{{{e^{ - i\phi }}\cos \theta \sin \theta }}{{\sqrt 2 }}}&{\frac{{{e^{ - 2i\phi }}{{\sin }^2}\theta }}{2}}\\
				{\frac{{{e^{i\phi }}\cos \theta \sin \theta }}{{\sqrt 2 }}}&{{{\sin }^2}\theta }&{ - \frac{{{e^{ - i\phi }}\cos \theta \sin \theta }}{{\sqrt 2 }}}\\
				{\frac{{{e^{2i\phi }}{{\sin }^2}\theta }}{2}}&{ - \frac{{{e^{i\phi }}\cos \theta \sin \theta }}{{\sqrt 2 }}}&{{{\cos }^2}\theta  + \frac{{{{\sin }^2}\theta }}{2} - \frac{{g{\mu _B}B}}{D}}
		\end{array}} \right)
	\end{array}
	\label{eq:hlab},
\end{equation}
\end{widetext}
where $D$ is the zero-field splitting, $R\left( t \right) = {R_z}\left( {\phi \left( t \right)} \right){R_y}\left( \theta  \right)$ is the rotation transformation, and ${R_j}\left( \theta \right) = \exp ( - i\theta{\bf{n}} \cdot {{\bf{S}}})$ for the rotation angle $\theta$ \rw{around the} $\bf{n}$ direction, $j = y,z$, and $\bf{S}$ is the spin operators. \rw{The Stark shift for NV centers induced by the electric field is	negligible and hence is not included in the equation. The Hamiltonian possesses three eigenstates ${\left| {{m_s},t} \right\rangle _{lab}}$ ($m_s = 0, \pm 1$). The detailed expressions can be found in the Supplementary Information. Based on its definition, the Berry phase can be calculated as \cite{Chudo2021}
\begin{equation}
	{\gamma _{{m_s}}} = i\int_0^t {_{lab}\left\langle {{m_s},t'} \right|\frac{\partial }{{\partial t'}}{{\left| {{m_s},t'} \right\rangle }_{lab}}dt'}  = {m_s}{\omega _r}t\cos \theta 
	\label{eq:berryphase}.
	\end{equation}
Here the Berry phase is calculated for an open-path and is hence gauge-dependent. The spin state of the NV center is observed through the interaction with a microwave magnetic field. In our experiment, the direction of the microwave is in the $yz$-plane and has a small angle $\theta' = 8.5 ^\circ$ relative to the $z$ axis, resulting from the asymmetric design of the waveguide. However, the dominant transition probability arises from the longitudinal ($z$) component. The expected value of the transition probability of the spin states interacting with the microwave can be expressed as
\begin{widetext}
\begin{equation}
	\begin{array}{l}
		_{lab}\left\langle { \pm 1,t} \right|{e^{i{H_{lab}}t/\hbar }}{e^{ - i{\gamma _{ \pm 1}}}}{H_{MW,z,lab}}{e^{i{\gamma _0}}}{e^{ - i{H_{lab}}t/\hbar }}{\left| {0,t} \right\rangle _{lab}}\\
		= \frac{1}{2}g{\mu _B}{B_{MW}}\cos \theta '{e^{i\left( { - {\omega _{MW}} + D \pm g{\mu _B}B\cos \theta  \mp {\omega _r}\cos \theta } \right)t}}_{lab}\left\langle { \pm 1,0} \right|{e^{i\theta {S_y}}}{S_z}{e^{ - i\theta {S_y}}}{\left| {0,0} \right\rangle _{lab}}
	\end{array}
	\label{eq:hmwzlab}.
\end{equation}
\end{widetext}
According to Eq.~\ref{eq:hmwzlab}, the transition of spin states from ${\left| {{m_s} = 0} \right\rangle _{lab}}$ to ${\left| {{m_s} =  \pm 1} \right\rangle _{lab}}$ can be driven by a microwave at the resonance frequency of $D \pm g{\mu _B}B\cos \theta \mp {\omega _r}\cos \theta$, where the frequency shift $\mp {\omega _r}\cos \theta$ is due to the Berry phase induced by the mechanical rotation.}

\rw{We first investigate the effect of the Berry phase in the absence of an external magnetic field. Fig.~\ref{fig:3}(a) shows the diagram of a NV center rotating around the $z$ axis without an external magnetic field.} To observe the frequency shift due to fast rotation, ODMR measurements of the levitated nanodiamond are carried out at different rotation frequencies. Fig.~\ref{fig:3}(b) displays ODMRs at rotation frequencies of 0.1 MHz (bule circles) and 14 MHz (red squares). The full width at half maximum (FWHM) of the ODMR at $\omega_r = 2\pi \times 14$ MHz is clearly larger than that at $\omega_r = 2\pi \times 0.1$ MHz \rw{, which} is caused by the \rw{Berry phase due to rotation}. The FWHM of the ODMR at different rotation frequencies is shown in \rw{Fig.~\ref{fig:3}(c)}. The blue circles are the experimental results. The red and orange curves are theoretical results for $\theta  = {0^ \circ }$ and $\theta  = {20^ \circ }$, \rw{respectively}. Experimentally, the NV ensemble contains NV centers with four orientations. \rw{Based on Eq.~\ref{eq:hmwzlab},} the broadening of the ODMR spectrum is mainly determined by NV centers with the smallest $\theta$, which \rw{have the largest frequency shift induced by the Berry phase}(Supplementary Fig. 6). \rw{The frequency shift of $\mp {\omega _r}\cos \theta$ is insensitive to the angle $\theta$ for small $\theta$. This explains why the theoretical  results for $\theta  = {0^ \circ }$ and $\theta  = {20^ \circ }$ are similar, and both agree well with the experimental results. All data shown in Fig.~\ref{fig:3}(b),(c) are taken from one levitated diamond.}

\begin{figure*}[th]
	\includegraphics[width=\textwidth]{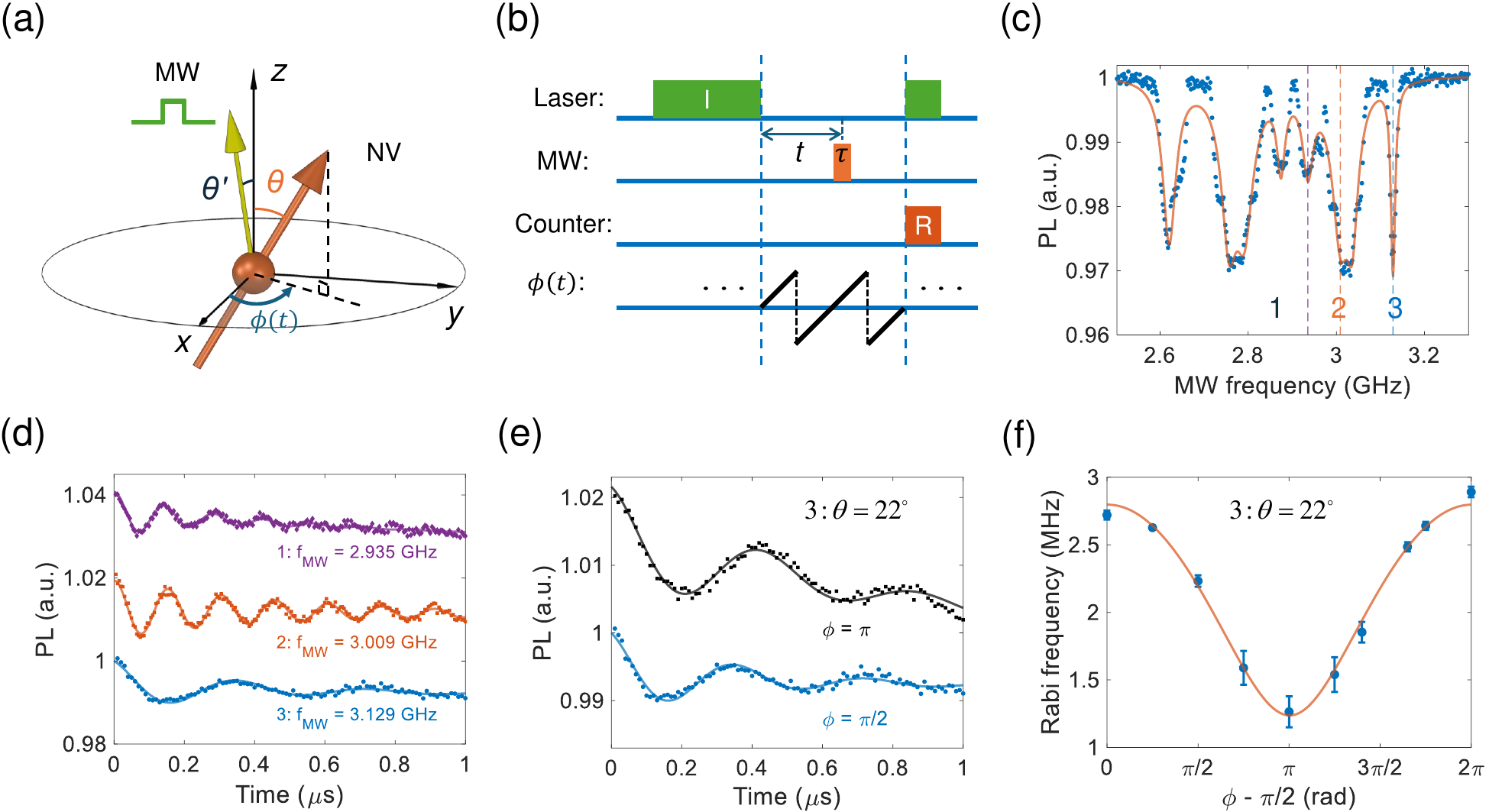}
	\caption{\label{fig:4} Quantum control of NV centers in a levitated nanodiamond in high vacuum with a rotation frequency of 100~kHz. (a) Schematic of the Rabi oscillation measurement at different rotation phase $\phi (t)$. The angle $\theta'$ between the magnetic component of microwave and the $z$ axis is $8.5 ^\circ$. (b) Pulse sequence of the Rabi oscillation measurement. (c) ODMR of the levitated nanodiamond. (d) Measured Rabi oscillations of NV centers at three different orientations. The Rabi frequencies are 7.10 MHz, 6.57 MHz and 2.80 MHz at the ODMR frequencies of 2.935 GHz, 3.009 GHz and 3.129 GHz, respectively. (e) Rabi oscillation of NV centers with $\theta = 22 ^\circ$ \rw{corresponding to the resonance frequency of 3.129 GHz (dip 3)}. The blue circles and \rw{black} squares are measured at rotation phase of \rw{$\phi = \pi/2$ and $\phi = \pi$}, respectively. (f) Rabi frequency at $\theta = 22 ^\circ$ (blue circles) as a function of the rotation phase $\phi$. The red curve is the theoretical prediction.}
\end{figure*}

\rw{To determine the frequency shift as a function of the rotational frequency unambiguously, an} external magnetic field can be applied to separate the energy levels of NV centers along four different orientations. 
\rw{Here} we apply a static magnetic field of about 100 G along the $z$ axis to separate energy levels (Fig. \ref{fig:3}\rw{(d)}). \rw{Data shown in Fig.~\ref{fig:3}(e),(f) are taken from one levitated diamond, which is different from the one used for Fig.~\ref{fig:3}(b),(c)}. In Fig. \ref{fig:3}\rw{(e)}, the red squares show the ODMR spectrum measured at a rotation frequency of 0.1 MHz. \rw{The linewidths of ODMR dips for levitated diamond NV centers are broader than those for fixed NV centers due to the continuous change of NV orientations relative to the magnetic field.} 	Compared with the ODMR spectrum of a levitated nanodiamond without stable rotation (gray circles), the linewidth of each dip \rw{for a diamond rotating at 0.1 MHz} is narrower. This clearly demonstrates that fast rotation can stabilize the orientation of the levitated nanodiamond. Now  we consider the NV centers with the smallest $\theta$ (largest Zeeman shift) and the transition between the state $\left| {{m_s} = 0} \right\rangle$ and the state $\left| {{m_s} =  + 1} \right\rangle$ as an example. The electron spin resonance frequency is 3.120 GHz at $\omega_r = 2\pi \times 0.1$ MHz for this transition. The corresponding angle between the NV axis and the rotation axis is $\theta = 20.7^\circ$, which is calculated based on the transition frequency and the magnitude of the external magnetic field. We then measure the resonance frequency at different \rw{clockwise (unless otherwise specified, all are viewed from the positive z direction)} rotation frequencies, as shown in Fig. \ref{fig:3}\rw{(f)}. The resonance frequency increases following the increase of the rotation frequency. The experimental data points are in between the theoretically calculated curves for $\theta=20.7^\circ$ (\rw{green} solid line) and \rw{$\theta = 24.0^\circ$ (violet} dashed line), indicating the orientation of the NV axes changes slightly when the rotation frequency increases. This is because the electric dipole moment of the levitated nanodiamond is not exactly perpendicular to the axis of the largest or the smallest moment of inertia. Once the rotation frequency increases, the nanodiamond tends to rotate along its stable axis and the driving torque is not large enough to keep its former orientation. \rw{The magenta dashed curve is a linear fitting of the resonance frequency. The orientation of the NV center can be calculated by the resonance frequency at the various rotation frequencies.  The angle $\theta$ changes by approximately $3.3^\circ$ at $\omega_r = 2\pi \times 10$ MHz, compared with that at $\omega_r = 2\pi \times 0.1$ MHz. A rotating diamond can also serve as a gyroscope \cite{Soshenko2021,Jarmola2021}.}

\rw{The effect of the Berry phase in a levitated nanodiamond rotating at the counterclockwise direction is shown in Supplementary Fig. 6. The resonance frequency between the state $\left| {{m_s} = 0} \right\rangle$ and the state $\left| {{m_s} =  + 1} \right\rangle$ decreases as the rotation frequency increases for counterclockwise rotation (Supplementary Fig. 6(c)), which is different from that of the levitated nanodiamond rotate clockwise (Fig. \ref{fig:3}(f)).}

\subsection{Quantum control of fast rotating NV centers}

\begin{figure*}
	\includegraphics[width=0.7\textwidth]{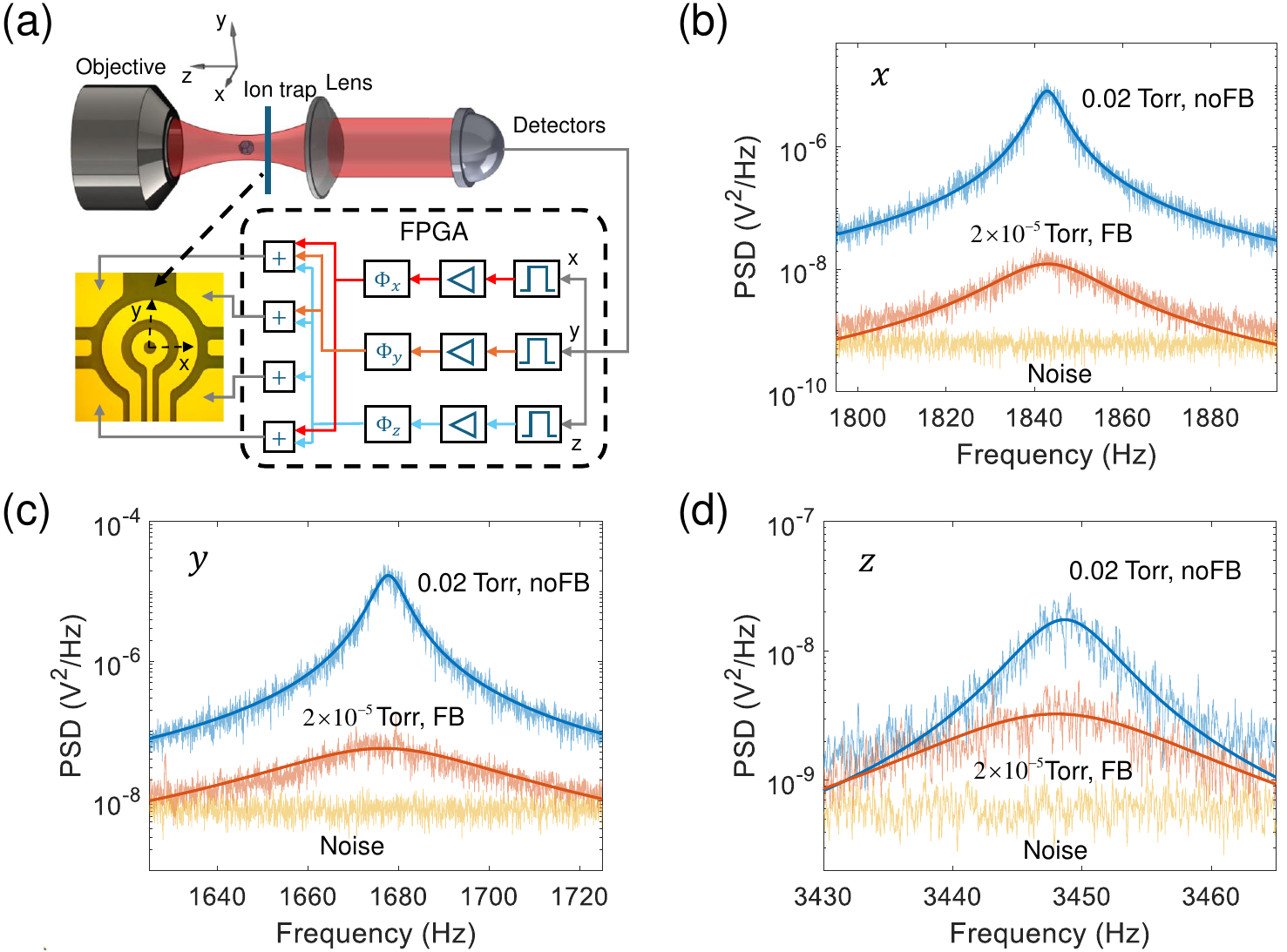}
	\caption{\rw{Feedback cooling of the CoM motion of a levitated nanodiamond in the ion trap. (a) Schematic diagram of the feedback cooling method. (b), (c), (d): PSDs of the CoM of the levitated nanodiamond along the (b) $x$, (c) $y$ and (d) $z$ directions at the pressure of 0.02 Torr without cooling (blue curves) and at the pressure of $2.0 \times 10^{-5}$ Torr with feedback cooling (red curves). The orange curves are the noise floors. Based on the fitting, the effective temperature of the CoM motion with feedback cooling are $1.2 \pm 0.3$ K, $3.5 \pm 0.4$ K and $86 \pm 26$ K along the $x$, $y$ and $z$ directions, respectively.}}
	\label{fig:5}
\end{figure*}

Quantum control of spins \rw{is} important for creating superposition states \cite{Yin2013,Scala2013,Rusconi2022} and performing advanced quantum sensing protocols \cite{RevModPhys.89.035002}. Here we apply a resonant microwave pulse to demonstrate quantum state control of fast rotating NV centers. The spin state can be read out by measuring the emission PL. Because a weak 532 nm laser is used to avoid significant heating, the initialization time should be long enough to prepare the NV spins to the $\left| {{m_s} = 0} \right\rangle$ state. When the laser intensity is 0.113 W/mm$^2$, the initialization time is 1.05 ms (Supplementary Fig. 7). This is shorter than the spin relaxation time (\rw{$T_1 \sim 3.6$ ms}) of this levitated nanodiamond (Supplementary Fig. 7). We also measure Rabi oscillation of a nanodiamond fixed on a glass cover slip \rw{with the same 532 nm laser intensity for comparison}. We get similar results for both high and low intensities of the 532 nm laser (Supplementary Fig. 8). Due to the $\Omega$-shape of the microwave antenna, the orientation of the magnetic field of the microwave is \rw{located in the $yz$-plane and} slightly different from the $z$ axis with an angle of about $\theta ' = 8.5 ^\circ$ (Fig. \ref{fig:4}(a)). \rw{So, ${{\bf{n}}_{MW}} = \left( { - \sin \theta ',0,\cos \theta '} \right)$}. The effective microwave magnetic field acting on NV spins\rw{, with the orientation of ${{\bf{n}}_{NV}} = \left( {\cos \phi \left( t \right)\sin \theta ,\sin \phi \left( t \right)\sin \theta ,\cos \theta } \right)$,} changes as a function of the rotation phase $\phi (t)$ of the levitated nanodiamond. \rw{The Rabi frequency $\Omega_{Rabi}$ can be written as \cite{Wood2020,Wood2021}}:
\begin{eqnarray}
	\Omega_{Rabi}  \propto \sqrt{1 - {\left( {\cos \theta \cos \theta ' - \sin \phi \left( t \right)\sin \theta \sin \theta '} \right)^2}}
	\label{eq5}.
\end{eqnarray}
\rw{Therefore,} it is necessary to synchronize the microwave pulse and the rotation phase of the levitated nanodiamond. Fig. \ref{fig:4}(b) shows the pulse sequence of the Rabi oscillation measurement. The time gap between the initialization and the readout laser pulses is twice of the rotation period, which allows us to apply the microwave pulse at an arbitrary rotation phase between 0 and $2\pi$.

The measured Rabi oscillations between the state $\left| {{m_s} = 0} \right\rangle$ and the state $\left| {{m_s} =  + 1} \right\rangle$ of NV centers are shown in Fig. \ref{fig:4}(d), (e). All these measurements are carried out at a rotation frequency of 100 kHz. The rotation period is 10 $\mu$s which is much longer than the microwave pulse. For NV centers with different orientations, the Rabi frequencies are different. The measured Rabi frequencies are 7.10 MHz, 6.57 MHz and 2.80 MHz when the applied microwave frequencies are 2.936 GHz (dip 1), 3.009 GHz (dip 2) and 3.129 GHz (dip 3), respectively (Fig. \ref{fig:4}(d)). Fig. \ref{fig:4}(e) shows Rabi oscillations of the NV centers with $\theta = 22 ^\circ$ at different rotation \rw{phases}. The blue circles and \rw{black} squares are measured at the rotation phase of \rw{$\phi = \pi/2$ and $\phi = \pi$, respectively.} The corresponding Rabi frequencies are 2.72 MHz and 2.23 MHz due to the different \rw{projections} of the microwave magnetic field along the NV axis. We also apply microwave pulse at other rotation phases to explore how it affects the Rabi frequency. Fig. \ref{fig:4}(f) shows the Rabi frequency $\Omega_{Rabi}$ for NV centers with $\theta = 22 ^\circ$ (blue circles) as a function of the rotation phase. The Rabi frequency is smallest at \rw{$\phi = 3\pi/2$}. The red curve is the theoretical prediction, which agrees well with our experimental results.

\subsection{\rw{Feedback cooling of the Center-of-Mass motion}}

\rw{To study quantum spin-mechanics and use a levitated diamond for precision measurements, it will be crucial to reduce the energy of the CoM motion of a levitated diamond. Our integrated Paul trap has multiple electrodes, which can be used for feedback cooling. Because of the high quality factor of the CoM in high vacuum and the low frequency of the CoM motion, we add $\pi/2$ phase delays to the position signals of the levitated diamond to obtain its velocity signals. We then apply  electric forces (on the charged diamond) proportional to the velocities but with opposite signs to cool the CoM motion. The schematic diagram is shown in Fig.~\ref{fig:5}(a). The feedback loop is implemented through an FPGA (Field Programmable Gate Array). The motion signals are read out, followed by band-pass filters, amplifiers and phase delayers, and then fed back to the four electrodes at the corners of the ion trap. Figs.~\ref{fig:5}(b)-(d) show the PSDs of the CoM motion of a levitated nanodiamond at the pressure of 0.02 Torr without feedback cooling (noFB, blue curves) and at the pressure of $2.0 \times 10^{-5}$ Torr with feedback cooling (FB, red curves). The orange curves are the corresponding noise floors. Based on the fitting, the final temperature of the CoM motion with feedback cooling are $1.2 \pm 0.3$ K, $3.5 \pm 0.4$ K and $86 \pm 26$ K along the $x$, $y$ and $z$ directions, respectively. The final temperatures are mainly limited by the small size of the center hole of the surface ion trap used for forward detection (Figs.~\ref{fig:5}(a)), which severely limited the NA of the detection system. The cooling efficiency can be improved in the future with backward detection  by using the backward scattered light of the levitated diamond collected by the objective lens. }

\section{Discussion}

In conclusion, we have levitated a nanodiamond at pressures below $10^{-5}$ Torr with a surface ion trap. We performed ODMR measurement of a levitated nanodiamond in high vacuum for the first time. The internal temperature of the levitated nanodiamond remains stable at about 350 K when the pressure is below $5 \times 10^{-5}$ Torr, which means stable levitation with an ion trap will not be limited by heating even in ultrahigh vacuum. This offers a unique platform for studying fundamental physics, such as massive quantum superposition \cite{Yin2013,Scala2013,Rusconi2022}. 

Additionally, we apply a rotating electric field that exerts a torque on the levitated nanodiamond to drive it to rotate at high speeds \rw{up to 20 MHz}. \rw{20 MHz rotation can generate a pseudo-magnetic field of 0.71 mT for an electron spin, and a pseudo-magnetic field of 6.5 T for an $^{14}$N nuclear spin.} With this method, the rotation frequency of a levitated nanodiamond is extremely stable and easily controllable. The \rw{effect of the Berry phase}  generated by rotation \cite{Chen2019} is observed with the embedded NV center electron spins. This will be useful for creating a gyroscope for rotation sensing \cite{Ledbetter2012,Soshenko2021,Jarmola2021}. We also demonstrate quantum  control of rotating NV centers in high vacuum, which will be important for using spins to create nonclassical states of mechanical motion \cite{Yin2013,Scala2013,Rusconi2022}. \rw{Using feedback cooling, the CoM of the levitated nanodiamond is cooled in all three directions with a minimum temperature of about 1.2 K along one direction.}

\rw{The maximum rotation frequency in this experiment is limited by the bandwidth of the multichannel waveform generation system for generating the phase-shifted signals on the four electrodes. The rotation frequency can be much higher with a better waveform generation system. Furthermore, in the presence of a DC external magnetic field, the NV centers within a rotating nanodiamond experience an AC magnetic field. Quantum sensing of an AC magnetic field can have a higher sensitivity compared to that of a DC magnetic field \cite{Wood2022}. Consequently, the mechanical rotation can enhance the sensitivity of a magnetometer in measuring DC magnetic fields. By using purer diamond particles, i.e. CVD diamonds, a higher excitation power of the 532 nm laser can be employed to reduce the initialization time of NV centers.}

\section*{Methods}

\subsection{Experiment setup and materials}

The surface ion trap is fabricated on a sapphire wafer by photolithography. The chip is fixed on a 3D stage and installed in a vacuum chamber. The AC high voltage signal used to \rw{levitate} nanoparticles and the microwave used for quantum control are combined with a bias tee to be delivered to the chip. A 532 nm laser beam is incident from the bottom to excite diamond NV centers. The photoluminescence (PL) is collected by an objective lens with a numerical aperture (NA) of 0.55. A 1064 nm laser beam focused by the same objective lens is used to monitor both the center-of-mass (CoM) motion and the rotation of the levitated nanoparticle. The PL is separated with the 532 nm laser and the 1064 nm laser by dichroic mirrors. The counting rate and optical spectrum of the PL are measured by a single photon couting module and a spectrometer. The processes of particle launching and trapping are monitored by two cameras.

\rw{The diamond particles were acquired from Adamas Nano. The product model is MDNV1umHi10mg (1 micron Carboxylated Red Fluorescence, 1 mg/mL in DI Water, $\sim$3.5 ppm NV). The experimental data shown in the main text of the manuscript are obtained from four different diamond particles. The data presented in Fig. \ref{fig:1}, Fig. \ref{fig:2}, and Figs. \ref{fig:3}(a-c) originate from measurements conducted on the same nanodiamond particle. Figs. \ref{fig:3}(d-f) show the data from a second nanodiamond particle, while the data in Fig. \ref{fig:4} is measured using the third nanodiamond particle. Fig. \ref{fig:5} uses the fourth diamond particle.}

\subsection{Internal temperature of a levitated nanodiamond}

In the experiment, we measure the ODMR of levitated nanodiamond NV centers to detect the internal temperature in the absence of an \rw{external} magnetic field. The zero-field Hamiltonian of NV center is: $H = DS_z^2\rw{/\hbar} + E\left( {S_x^2 - S_y^2} \right)\rw{/\hbar}$, where $D$ is the zero-field energy splitting between the states of $\left| {{m_s} =  0} \right\rangle$ and $\left| {{m_s} =  \pm 1} \right\rangle$, $E$ is the splitting between the states due to the strain effect. \rw{The small splitting between two dips in the ODMR spectra (Fig. 1(e)) without an external magnetic field is due to the $E$ term from strain in the nanodiamond}. The zero-field splitting $D$ is dependent on temperature \cite{Toyli2012PhysRevX,Hoang2016}:
\begin{eqnarray}
	D = {c_0} + {c_1}T + {c_2}{T^2} + {c_3}{T^3} + {\Delta _{pressure}} + {\Delta _{strain}}
	\label{eq6},
\end{eqnarray}
where $c_0 = 2.8697$ GHz, $c_1 = 9.7 \times 10^{-5}$ GHz/K, $c_2 = -3.7 \times 10^{-7}$ GHz/K$^2$, $c_3 = 1.7 \times 10^{-10}$ GHz/K$^3$, $\Delta _{pressure} = 1.5 \times 10^{-6}$ GHz/bar, and $\Delta _{strain}$ is caused by the internal strain effect. $\Delta _{pressure}$ is smaller and can be neglected in vacuum. Fig. \ref{fig:1}(e) is the ODMR measured at the pressure of 10 Torr (blue circles) and $6.9 \times 10^{-6}$ Torr (red squares). The zero-field splitting obtained by fitting can be used to calculate the temperature of the levitated nanodiamond.

The internal temperature $T$ of a levitated nanodiamond is determined by the balance between heating and cooling effects \cite{Liu2006,Chang2010}: 
\begin{eqnarray}
	{A_a} = {A_{gas}}p\left( {T - {T_0}} \right) + {A_{bb}}\left( {{T^5} - T_0^5} \right)
	\label{eq1},
\end{eqnarray}
where $A_a = \sum\nolimits_\lambda  {{\eta _\lambda }{I_\lambda }V}$ is the heating of the excitation laser ($\lambda$ = 532 nm) and the detecting laser ($\lambda$ = 1064 nm), $\eta _\lambda$ is the absorption coefficient of nanodiamond and $I_\lambda$ is the laser intensity, $V$ is the volume of nanodiamond. The first term at the right side of the equation is the cooling rate caused by gas molecule collisions, $A_{gas} = \frac{1}{2}\kappa \pi {R^2}v{T_0}\frac{{\gamma' + 1}}{{\gamma' - 1}}$, $\kappa \approx 1$ is the thermal accommodation coefficient, $R$ is the radius of nanodiamond, $v$ is the mean thermal speed of gas molecules, $\gamma'$ is the specific heat ratio ($\gamma' = 7/5$ for air near room temperature), $p$ is the pressure, $T_0$ is the thermal temperature. The last term is the cooling rate of black-body radiation. $A_{bb} = 72\zeta \left( 5 \right)Vk_B^5/\left( {{\pi ^2}{c^3}{\hbar ^4}} \right){\mathop{\rm Im}\nolimits} \left( {\frac{{\varepsilon  - 1}}{{\varepsilon  + 2}}} \right)$, where $\zeta \left( 5 \right) \approx 1.04$ is the Riemann zeta function, $k_B$ is the Boltzmann constant, $c$ is the vacuum light speed, $\hbar$ is the reduced Planck's constant, $\varepsilon$ is a constant and time-independent permittivity of nanodiamond across the black-body radiation spectrum. By measuring the internal temperature as a function of the intensities of the 532 nm laser and the 1064 nm laser, the absorption coefficients of the nanodiamond are estimated to be 111~cm$^{-1}$ at 532 nm and 5.87~cm$^{-1}$ at 1064 nm (Supplementary Fig. 3).

\begin{acknowledgments}
	We thank Jun Ye for helpful discussions. We acknowledge the support from the National Science Foundation under Grant PHY-2110591 and the Office of Naval Research under Grant No. N00014-18-1-2371. This project is also partially supported by the Laboratory Directed Research and Development program at Sandia National Laboratories, a multimission laboratory managed and operated by National Technology and Engineering Solutions of Sandia LLC, a wholly owned subsidiary of Honeywell International Inc., for the U.S. Department of Energy’s National Nuclear Security Administration under Contract No. DE-NA0003525. This paper describes objective technical results and analysis. Any subjective views or opinions that might be expressed in the paper do not necessarily represent the views of the U.S. Department of Energy or the United States Government.
\end{acknowledgments}

%\section*{Date availability}
%\rw{Source data for figures in the main text are provided with this paper. Other data that support the findings of this study are available from the corresponding author upon reasonable request.}

%\section{References}

%\bibliography{references}% Produces the bibliography via BibTeX.

%\section*{Author contributions}
%T.L., Y.J., K.S. and P.J. conceived and designed the project. Y.J. and K.S. built the setup. Y.J. performed measurements and calculations. Y.J., K.S., T.L., X.G., C.Z., and A.J.G. discussed the results. T.L. supervised the project. All authors contributed to the writing of the manuscript.

%\section*{Competing interests}
%The authors declare no competing interests.

%\vspace{12pt}
%\textbf{Correspondence} and requests for materials should be addressed to Tongcang Li.
%\nocite{*}

\newpage
\onecolumngrid
\appendix
\renewcommand{\figurename}{Supplemental Fig.}
\setcounter{figure}{0} 
\section*{Supplementary Information}

\section{Experiment setup and surface ion trap}

\begin{figure*}[bp]
	\includegraphics[width=\textwidth]{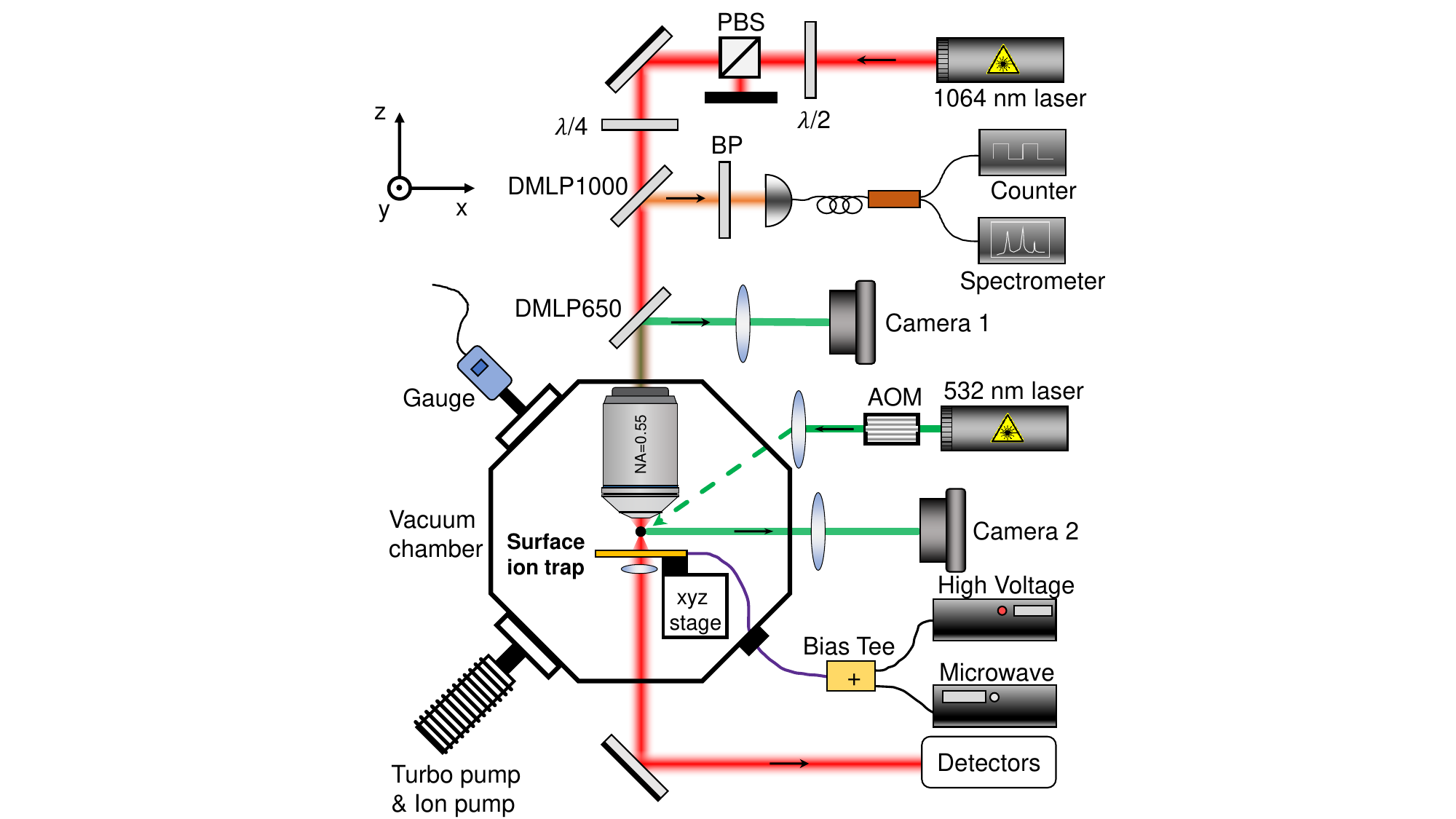}
	\caption{\label{fig:s1} Schematic diagram of the experimental setup. The surface ion trap is fixed on a 3D stage and installed in the vacuum chamber.  A 532 nm laser beam controlled by an acousto-optic modulator (AOM) is incident from the bottom to excite diamond NV centers in a levitated nanodiamond. The photoluminescence (PL) is collected by an objective lens. A 1064 nm laser beam focused by the same objective lens is used to monitor the motion of levitated nanodiamonds. The PL is separated with the 532 nm laser and the 1064 nm laser by two long-pass dichroic mirrors (DMLP650 and DMLP1000). PBS: Polarizing beam splitter; $\lambda/2$: half-wave plate; $\lambda/4$: quarter-wave plate; BP: band-pass filters}
\end{figure*}

Supplementary Fig. \ref{fig:s1} shows the schematic diagram of our experimental setup. The surface ion trap is fabricated on a sapphire wafer, which has a high transmittance for visible and near infrared laser. The chip is fixed on a 3D stage and installed in a vacuum chamber. The AC high voltage signal used to \rw{levitate} nanodiamonds and the microwave used for quantum control are combined with a bias tee and delivered to the chip. A 532 nm laser beam is incident from the bottom to polarize a levitated nanodiamond. The photoluminescence (PL) is collected by an objective lens with a numerical aperture (NA) of 0.55. A 1064 nm laser beam focused by the same objective lens is used to monitor the center-of-mass (CoM) motion and the rotational motion of levitated nanodiamonds. The PL is separated with the 532 nm laser and the 1064 nm laser by two long-pass dichroic mirrors. The count rate and the optical spectrum of the PL are measured by a photon counter and a spectrometer. There are two cameras that monitor the procedure of particles loading and the trapping position of levitated nanoparticles.

\begin{figure*}
	\includegraphics[width=\textwidth]{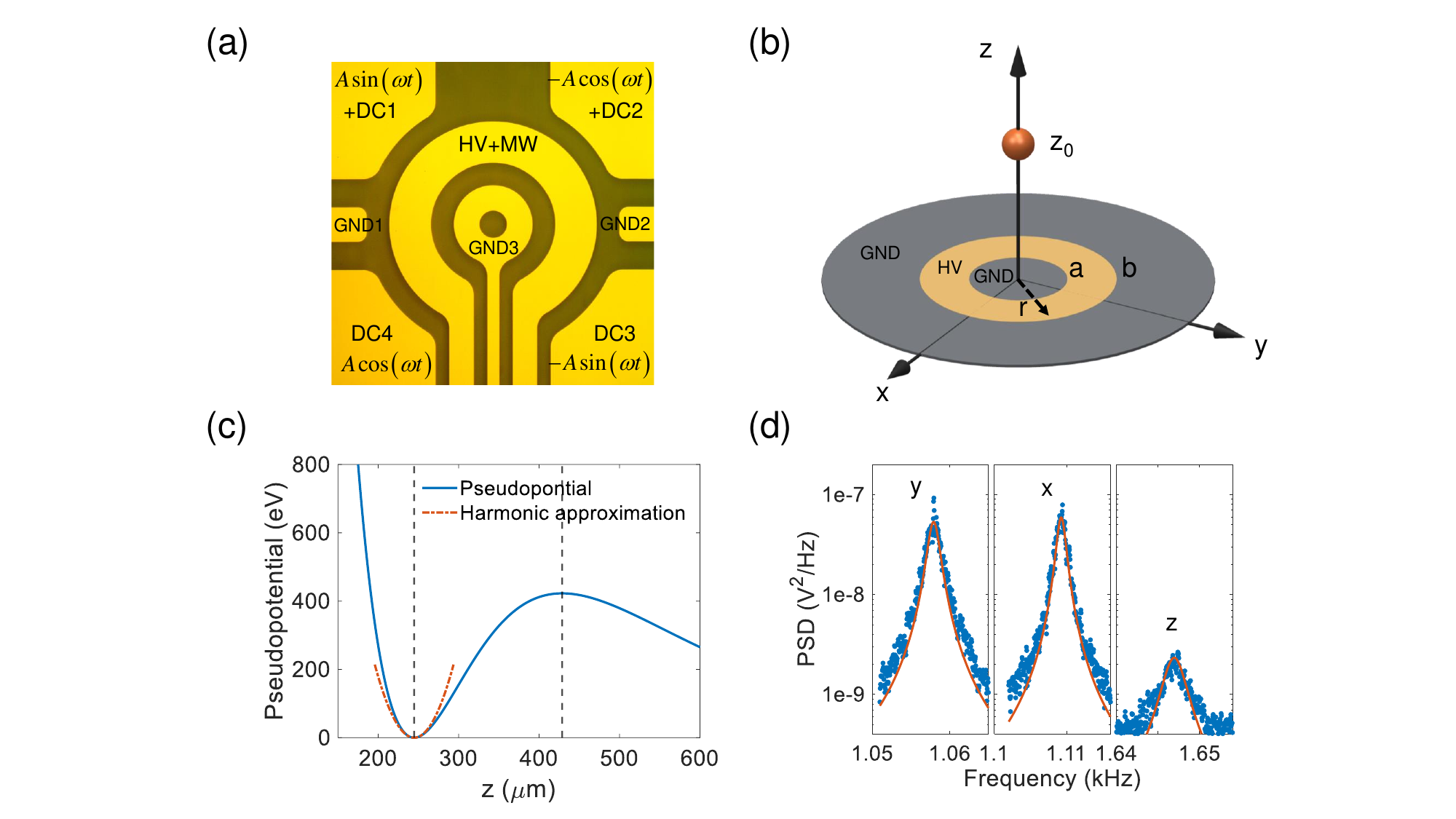}
	\caption{\label{fig:s2} Surface ion trap and CoM motion of a levitated nanodiamond. (a) Optical image of the surface ion trap. The outer four electrodes are used for compensation of residual charges on the chip surface. Trapping voltage and microwave can be delivered by a bias-tee simultaneously. (b) Equivalent surface ion trap with concentric rings design. Ring area $a \le r \le b$ is connected to an AC high voltage and the remaining parts are grounded. (c) Theoretical pseudopotential calculation in $z$ axis of a concentric rings ion trap. The trapping position is located at $z_0 = 245$ $\mu$m and the trapping depth is around 420 eV. The parameters used for this calculation are listed in Table \ref{tab:1}. (d) PSDs of the CoM motion of the levitated nanodiamond in three dimensions at the pressure of 0.01 Torr. The radius is about 264 nm by the Lorentzian fitting.}
\end{figure*}

Instead of a surface ring ion trap with concentric rings \cite{Kim2010,Li2017}, we use an $\Omega$-shaped circuit to deliver both a high voltage for trapping and a microwave for controlling NV centers. A blank hole in the center allows the probing laser to travel through. The planar design conveniently provides six-directional detection. To theoretically calculate the trapping parameters, including the trapping position and the trapping depth, we approximate the homemade surface ion trap (Supplement Fig. \ref{fig:s2}(a)) as a perfect ring ion trap (Supplement Fig. \ref{fig:s2}(b)). The area $a \le r \le b$ is connected to an AC high voltage driving signal with a frequency of $f_d$ and an amplitude of $V_d$, while the remaining parts are grounded. For a surface ion trap, the motion of particles in $z$ axis (perpendicular to the chip surface) is more critical. The motion equation of a nanoparticle along the $z$ axis can be approximately written as \cite{Kim2010}:
\begin{eqnarray}
	m\frac{{{d^2}z}}{{d{t^2}}} =  - Q V_d\cos \left( 2\pi f_d t \right)f\left( {a,b} \right)\left( z - z_0 \right)
	\label{eq:s1},
\end{eqnarray}
where $m$ is the mass of the nanoparticle, $Q$ is the charges, $z_0 = \sqrt {{a^{4/3}}{b^{4/3}}/\left( {{a^{2/3}} + {b^{2/3}}} \right)}$ is the trapping position located at the zero field point, and $f(a,b)$ is the geometric factor given by
\begin{eqnarray}
	f\left( {a,b} \right) = \sqrt {\frac{{9{{\left( {{b^{2/3}} - {a^{2/3}}} \right)}^2}{{\left( {{b^{2/3}} + {a^{2/3}}} \right)}^6}}}{{{a^{4/3}}{b^{4/3}}{{\left( {{a^{4/3}} + {a^{2/3}}{b^{2/3}} + {b^{4/3}}} \right)}^5}}}}
	\label{eq:s2}.
\end{eqnarray}
Eq. \ref{eq:s1} \rw{shares} the same forms as Mathieu equation. The trapping eigenfrequency along z direction can be solved as:
\begin{eqnarray}
	\omega _z = \frac{q}{{2\sqrt 2 }}2\pi f_d = \frac{{Q{V_d}}}{{2\sqrt 2 \pi m{f_d}}}f\left( {a,b} \right)
	\label{eq:s3}.
\end{eqnarray}

For a regime with small displacement of a levitated nanodiamond, the electric potential can be approximated as a harmonic potential near the trapping region:
\begin{eqnarray}
	V_{potential}\left( z \right) = \frac{1}{2}m\omega _z^2{\left( {z - {z_0}} \right)^2} = \frac{{{Q^2}V_d^2}}{{16{\pi ^2}mf_d^2}}{f^2}\left( {a,b} \right){\left( {z - {z_0}} \right)^2}
	\label{eq:s4}.
\end{eqnarray}
Generally, when the levitated nanodiamond moves away from the harmonic region, the pseudopotential can be written as \cite{Kim2010}:
\begin{eqnarray}
	V_{potential}\left( z \right) = \frac{{{Q^2}V_d^2}}{{16{\pi ^2}mf_d^2}}{\left| {\nabla \left( {\frac{1}{{\sqrt {1 + {{\left( {a/z} \right)}^2}} }} - \frac{1}{{\sqrt {1 + {{\left( {b/z} \right)}^2}} }}} \right)} \right|^2}
	\label{eq:s5}.
\end{eqnarray}

\begin{table}
	\caption{\label{tab:1} Parameters for the pseudopotential calculation of equivalent surface ion trap. $q_z$ satisfies the condition of stable 3D trapping.}
	\begin{ruledtabular}
		\begin{tabular}{cccccccc}
			Q/e & R/nm & $\rho$/(kg/$m^3$) & $V_d$/V & $f_d$/Hz & a/$\mu$m & b/$\mu$m & $q_z$\\
			\hline
			2000 & 264 & 3500 & 300 & $2\pi\times1.6\times10^4$ & 270 & 450 & 0.29\\
		\end{tabular}
	\end{ruledtabular}
\end{table}

The dimension of the surface ion trap is designed as $a=270 \mu$m and $b=450 \mu$m. We theoretically calculate the trapping potential of a levitated nanodiamond in $z$ axis, as shown in Supplement Fig. \ref{fig:s2}(c). The red dash-dotted curve and blue solid curve are calculated by Eq. \ref{eq:s4} and Eq. \ref{eq:s5}, respectively. All the parameters are summarized in Table \ref{tab:1}. The theoretical trapping position $z_0$ is 245 $\mu$m, which is very close to the simulation result 253 $\mu$m for the current ion trap design. The difference is due to the asymmetric ion trap design.

The trapping potential is dependent on the eigenfrequency of a levitated particle, which is proportional to the charge to mass ratio ($Q/m$). Thus, it is necessary to increase the charge number carried on particles to achieve stable levitation in an ion trap. In our experiment, \rw{the diamond particles were purchased from Adamas Nano and the product model is MDNV1umHi10mg (1 micron Carboxylated Red Fluorescence, 1 mg/mL in DI Water, ~3.5 ppm NV). These particles exhibit an average size of 750 nm. They are created by irradiating 2-3 MeV electrons on diamonds manufactured by static high-pressure, high-temperature (HPHT) synthesis and containing about 100 ppm of substitutional N.} The nanodiamonds are first sprayed out by  electrospray, which is supplied by a DC high voltage ($\sim 2kV$). Then the nanodiamonds are delivered to the trapping region of the surface ion trap with an extra linear Paul trap. 

After a nanodiamond is trapped, we apply a 1064 nm laser to measure the CoM motion of the levitated nanodiamond. Supplementary Fig. \ref{fig:s2}(d) is the PSDs of the CoM motion in x,y,z directions at the pressure of 0.01 Torr. The radius of the levitated nanodiamond is obtained to be about 264 nm based on the fitting of the PSDs. The experimental trapping frequency in z direction is about $\omega_z / 2 \pi = 1.65$ kHz. Using Eq.\ref{eq:s3}, the charge number is estimated to be about 2000 for this nanodiamond.  The surface ion trap creates an extremely deep potential well of 420 eV (Supplementary Fig. \ref{fig:s2}(c)). \rw{According to our experimental results}, the charge number of different levitated nanodiamonds varies from 1,000 to 10,000.

\section{Internal temperature of a levitated nanodiamond}

\begin{figure*}
	\includegraphics[width=\textwidth]{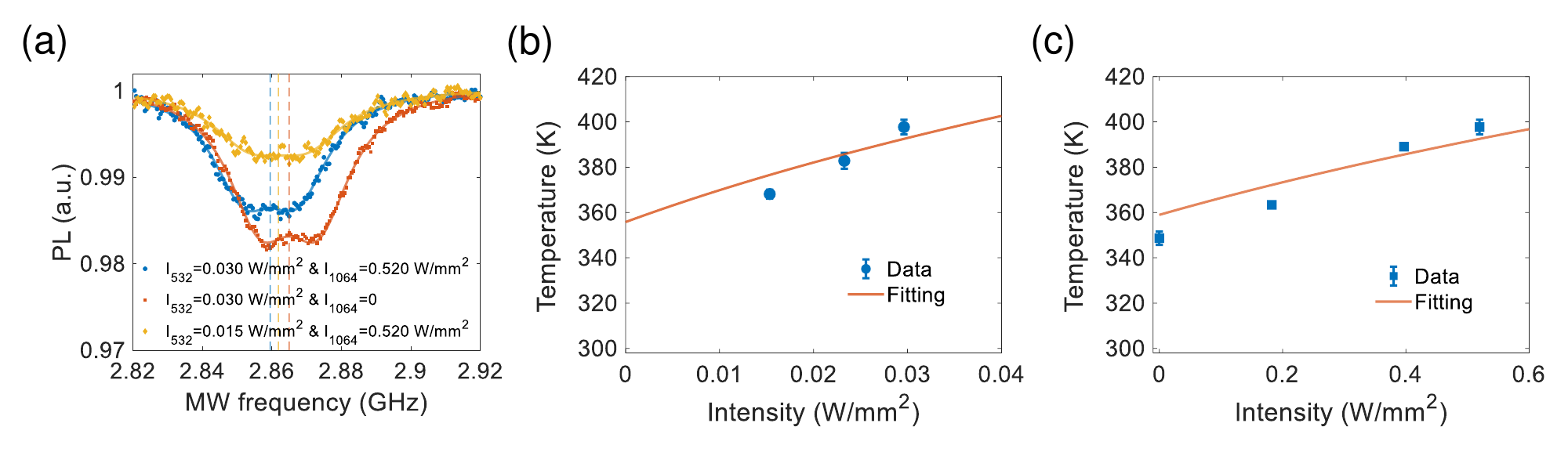}
	\caption{\label{fig:s3} Internal temperature of a levitated nanodiamond. (a) ODMRs of the levitated nanodiamond at different intensities of the 532 nm laser and the 1064 nm laser at the pressure of $1.3 \times 10^{-5}$ Torr. (b) Internal temperature as a function of the intensity of the 532 nm laser. The intensity of the 1064 nm laser is 0.520 W/mm$^2$. (c) Internal temperature as a function of the intensity of the 1064 nm laser. The intensity of the 532 nm laser is 0.030 W/mm$^2$.}
\end{figure*}

The heating by the 532 nm laser and the 1064 nm laser affects the stability of the levitated nanodiamond in vacuum. Here we measure the ODMRs of the levitated nanodiamond at different intensities of the 532 nm laser ($I_{532}$) and the 1064 nm laser ($I_{1064}$), as shown in Supplementary Fig. \ref{fig:s3}(a). The pressure is fixed to $1.3 \times 10^{-5}$ Torr. First, we adjust the intensity of the 532 nm laser from 0.015 W/mm$^2$ to 0.03 W/mm$^2$, while keeping $I_{1064} = 0.520$ W/mm$^2$. The internal temperature of the levitated nanodiamond is shown in Supplementary Fig. \ref{fig:s3}(b). Then we measure the temperature when the intensity of the 1064 nm laser is changed from 0 to 0.520 W/mm$^2$ while the intensity of the 532 nm laser is fixed at $I_{532} = 0.03$ W/mm$^2$ (Supplementary Fig. \ref{fig:s3}(c)). The red curves are the fittings by \cite{Liu2006,Chang2010}:
\begin{eqnarray}
	{A_a} = {A_{gas}}p\left( {T - {T_0}} \right) + {A_{bb}}\left( {{T^5} - T_0^5} \right)
	\label{eq:s6},
\end{eqnarray}
where the first term $A_a = \sum\nolimits_\lambda  {{\eta _\lambda }{I_\lambda }V}$ is the heating of the excitation laser ($\lambda$ = 532 nm) and the probe laser ($\lambda$ = 1064 nm), $\eta _\lambda$ is the absorption coefficient of nanodiamond and $I_\lambda$ is the laser intensity, $V$ is the volume of nanodiamond. The second term is the cooling rate caused by gas molecule collisions, $A_{gas} = \frac{{\kappa \pi {R^2}v}}{{2{T_0}}}\frac{{\gamma ' + 1}}{{\gamma ' - 1}}$, $\kappa \approx 1$ is the thermal accommodation coefficient of diamond, $R = 332$ nm is the radius of this nanodiamond (several different nanodiamonds are used in this experiment), $v$ is the mean thermal speed of gas molecules, $\gamma'$ is the specific heat ratio ($\gamma' = 7/5$ for air near room temperature), $p$ is the pressure, $T_0$ is the thermal temperature (298 K). Plug in the parameters, we get the coefficient $A_{gas}$ to be $1.74 \times {10^{ - 12}}$ m$^3 \cdot$s$^{ - 1} \cdot $K$^{-1}$ for this nanodiamond. The last term is for cooling due to black-body radiation, where $A_{bb} = 72\zeta \left( 5 \right)Vk_B^5/\left( {{\pi ^2}{c^3}{\hbar ^4}} \right){\mathop{\rm Im}\nolimits} \left( {\frac{{\varepsilon  - 1}}{{\varepsilon  + 2}}} \right)$, $\zeta \left( 5 \right) \approx 1.04$ is the Riemann zeta function, $k_B$ is the Boltzmann constant, $c$ is the vacuum light speed, $\hbar$ is the reduced Planck's constant, $\varepsilon$ is a constant and time-independent permittivity of nanodiamond across the black-body radiation spectrum. Based on the coefficient $A_{gas}$, we can calculate the absorption coefficients of 532 nm laser and 1064 nm laser to be 111 cm$^{-1}$ and 5.87 cm$^{-1}$, respectively.

\section{Rotation of a levitated nanodiamond driven by a rotating electric field}

\begin{figure*}
	\includegraphics[width=\textwidth]{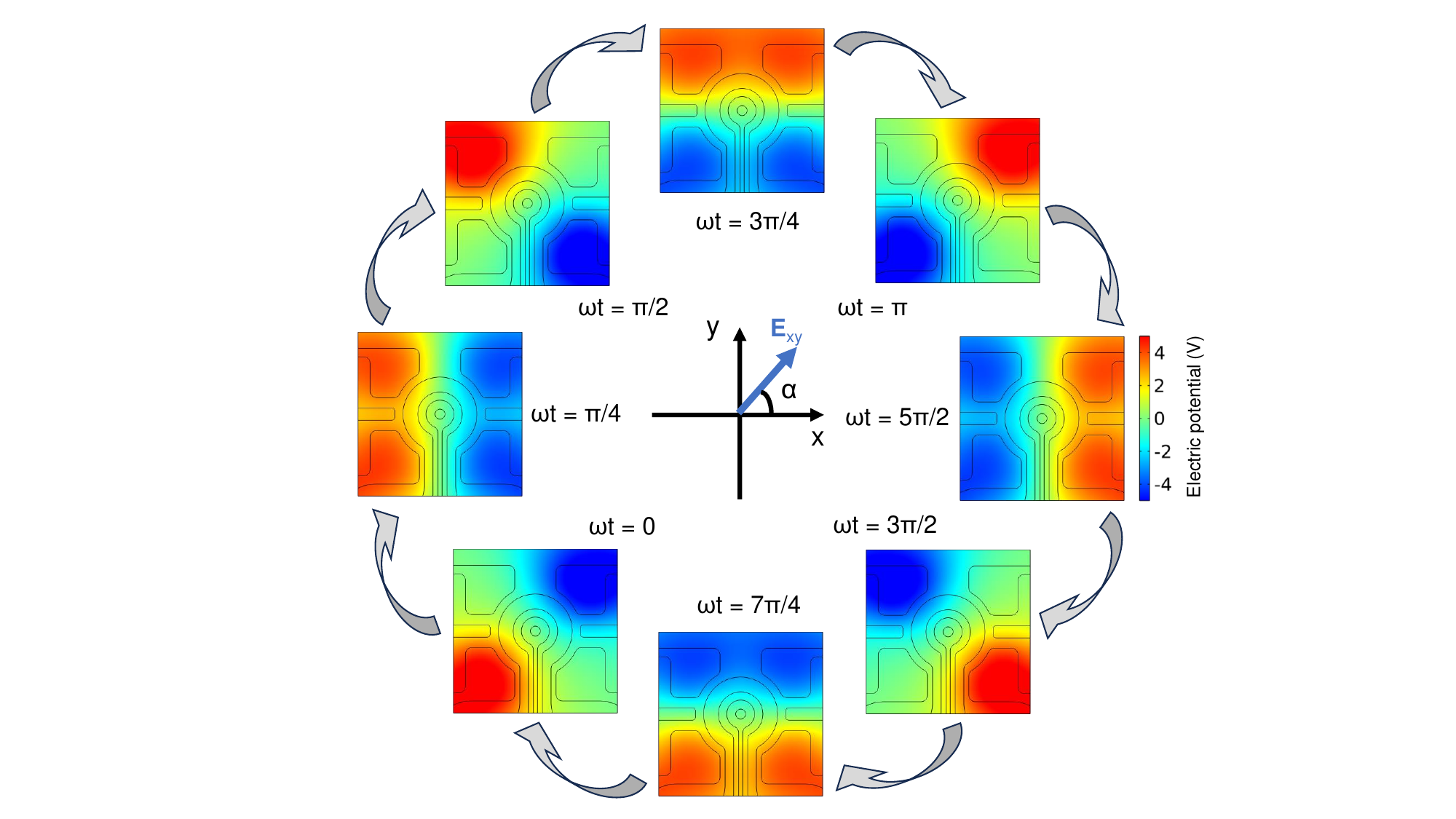}
	\caption{\label{fig:s4} Simulation of the rotating electric potential in $xy$-plane generated by the four electrodes. The rotation phase ($\omega t$) of each figure is changed from 0 to $2\pi$ by the step of $\pi/4$, and the direction of corresponding electric field rotate following the rotation phase.}
\end{figure*}

\begin{figure*}
	\includegraphics[width=\textwidth]{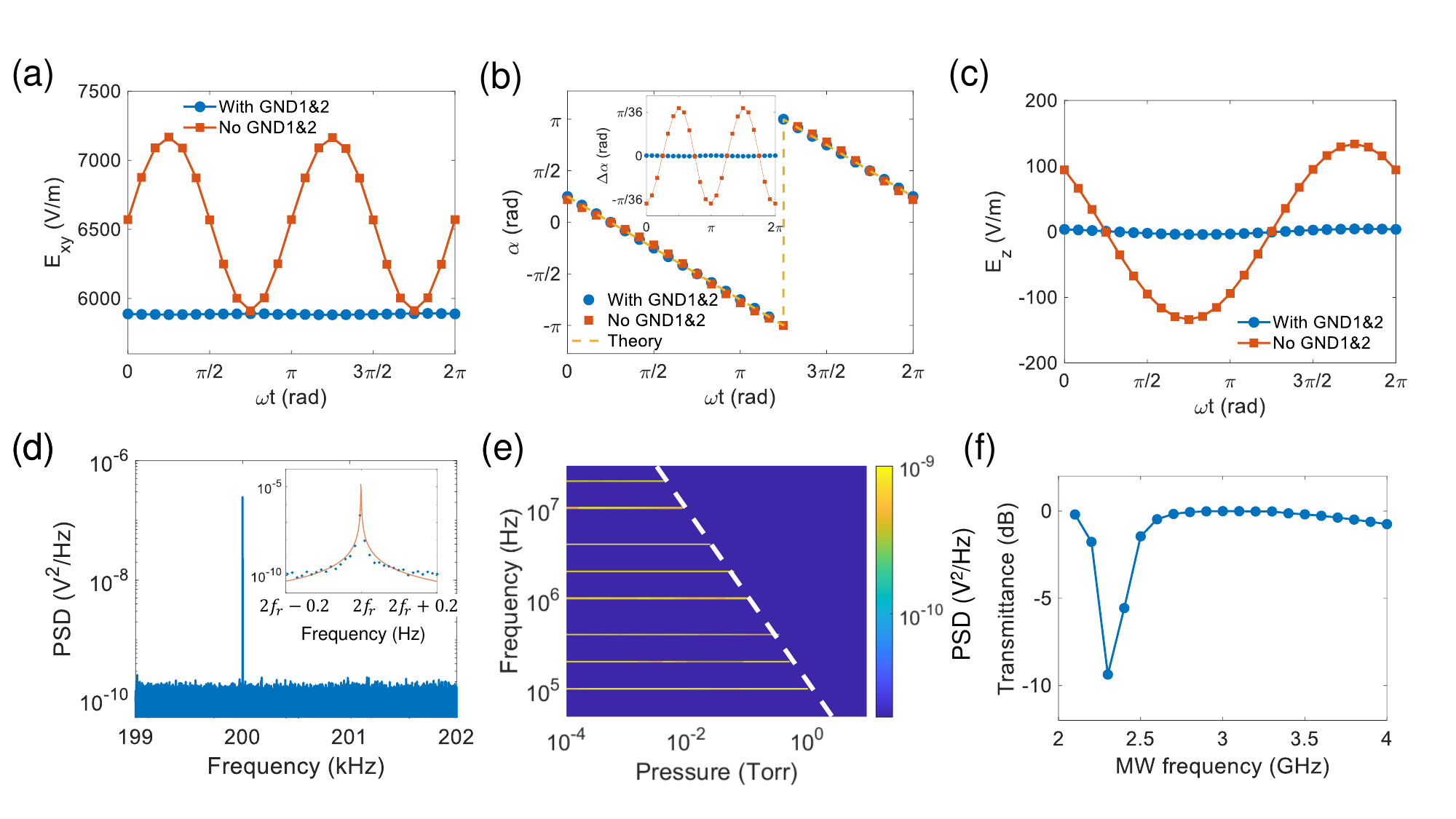}
	\caption{\label{fig:s5} Simulation of the rotating electric field and the rotational motion of a levitated nanodiamond. Time evolution of the $xy$-plane component (a) and direction (b), $z$ component (c) of the electric field with (blue circles) and without (red squares) the compensation electrodes of GND1 and GND2. $\alpha$ describes the orientation of the electric field ($\alpha$ = 0 indicates the electric field points to positive $x$ direction). (d) PSD of the rotational motion at the rotation frequency of 0.1 MHz. The linewidth is $9.9 \times 10^{-5}$ Hz by the fitting (inset). The ratio of the center frequency to the linewidth is $2 \times 10^9$. (e) PSDs of the rotational motion of the levitated nanodiamond at different pressures, showing the maximum rotation frequency at different pressures. The rotation frequencies are 0.05 MHz, 0.1 MHz, 0.2 MHz, 0.5 MHz, 1 MHz, 2 MHz, 5 MHz and 10 MHz, respectively. The upper limit of the rotation frequency is inversely proportional to pressure with the electric field driving. Particles stop rotating above the dashed white line. (f) Simulation of microwave transmittance of the surface ion trap.}
\end{figure*}

In our experiment, we use a rotating electric field to drive a levitated nanodiamond to rotate at high speed. The four electrodes at the corners are applied with 4 AC signals with the same frequency and amplitude but $\pi/2$ phase difference to generate a rotating electric field. The two grounded electrodes labeled by GND1 and GND2 (Supplementary Fig. 2(a)) are introduced to cancel the $z$ component of the rotating electric field and make the electric field more symmetric. \rw{We simulate the electric fields for both the trapping potential and the rotating field using the COMSOL software.} The simulation of the electric potential in $xy$-plane at different rotation phases are shown in Supplementary Fig. \ref{fig:s4}. The rotation phases are changed from 0 to $2\pi$ by steps of $\pi/4$ in the simulation. The dipole moment ($\bf{p}$) of a levitated nanodiamond is aligned to the direction of the electric field (${\bf{E}}_{xy}$) by the torque
\begin{eqnarray}
	{{\bf{M}}_{electric}} = {\bf{p}} \times {{\bf{E}}_{xy}} = \left| {\bf{p}} \right|\left| {{{\bf{E}}_{xy}}} \right|\sin \beta \cdot {\bf{z}}
	\label{eq:s7},
\end{eqnarray}
where $\beta$ is the angle between the dipole moment and the electric field, $\bf{z}$ is the unit vector along z direction. 

Based on the simulation as shown in Supplementary Fig. \ref{fig:s4}, the amplitude and the direction ($\alpha$) of the electric field in the $xy$-plane can be calculated, which is displayed in Supplementary Fig. \ref{fig:s5}(a) and \ref{fig:s5}(b). $\alpha$ = 0 indicates that the rotating electric field points to the positive $x$ direction. The blue circles and red squares are the simulations with and without the compensation electrodes GND 1 and GND 2. Ideally, the direction of the rotating electric field should be \rw{$\alpha \left( t \right) = \pi /4 - \omega t$}(orange dashed curve). The inset of Supplementary Fig. \ref{fig:s5}(b) is the asynchrony between the simulation result and an ideal rotation field with and without the compensation electrodes. The orientation of the electric field does not perfectly rotate at a constant speed in one period without the compensation electrodes (blue circles). The maximum deviation is $5.5^\circ$. It hurts the stability of the rotational motion of the levitated nanodiamonds and expand the linewidth of nanodiamond's rotation signal. Moreover, the $E_z$ component of the rotating electrical field oscillates with a large amplitude if there are no compensation electrodes (Supplementary Fig. \ref{fig:s5}(c)). The electrical field drives a levitated nanodiamond to oscillate in the $z$ direction, causing the loss of the levitated nanodiamond in high vacuum. The two compensation electrodes effectively solve these issues. The transmittance of a microwave through the $\Omega$-shaped circuit is simulated (Supplementary Fig. \ref{fig:s5}(f)) to ensure the microwave has low loss for frequencies from 2.6 GHz to 3.1 GHz.

Then we drive a levitated nanodiamond to rotate using the rotating electric field. The PSD of the rotational motion at the rotation frequency of 0.1 MHz is shown in Supplementary Fig. \ref{fig:s5}(d). The linewidth of the rotation signal is about $9.9 \times 10^{-5}$ Hz based on a Lorentzian fitting (inset of Supplementary Fig. \ref{fig:s5}(d)). Thus, the ratio of the center frequency to the linewidth is $2 \times 10^9$, demonstrating the rotational motion is ultra-stable with easy control by this method. 

Meanwhile, the rotational motion of the levitated nanodiamond is damped by the interaction with the remaining gas molecules in vacuum chamber. The damping torque of a sphere is \cite{Jin2021,Fremerey1982}
\begin{eqnarray}
	{M_{gas}} =  - I{\omega _r}{\gamma _d}
	\label{eq:s8},
\end{eqnarray}
where $I$ is the moment of inertia of the nanodiamond, $\omega _r$ is the angular velocity, ${\gamma _d} =40 \eta' p{R^2}/ 3mv$ is the damping rate of rotational motion, $\eta ' \approx 1$ is the accommodation factor accounting for the efficiency of the angular momentum transferred onto the nanodiamond by gas molecule collisions. Thus, the rotational motion equation can be written as:
\begin{eqnarray}
	I\frac{{d{\omega _r}}}{{dt}} = {M_{electric}} + {M_{gas}}
	\label{eq:s9}.
\end{eqnarray}
The maximum rotation frequency of the levitated nanodiamonds is obtained at $M_{electric} = - M_{gas}$ and $\beta = \pi/2$, which is limited by the pressure in the vacuum chamber. The maximum rotation frequency at a certain pressure is:
\begin{eqnarray}
	{\omega _{r\max }} = \left| {\bf{p}} \right|\left| {{{\bf{E}}_{xy}}} \right|/I{\gamma _d}
	\label{eq:s10}.
\end{eqnarray}

We measure the upper limit of the rotation frequency at different pressures (Supplementary Fig. \ref{fig:s5}(e)). The PSDs as functions of air pressure are measured at the rotation frequencies of 0.05 MHz, 0.1 MHz, 0.2 MHz, 0.5 MHz, 1 MHz, 2 MHz, 5 MHz and 10 MHz. The levitated nanodiamond stops rotating when the pressure is too large for that rotation frequency. The maximum rotation frequency is inversely proportional to the pressure (white dashed curve). The dipole moment of the nanodiamond ($R = 264$ nm) is estimated to be about $\left| {\bf{p}} \right| = 3.13 \times {10^{ - 25}}$ C$\cdot$m (1.96 e$\cdot \mu$m). We can adjust and lock the rotation of the levitated nanodiamond at arbitrary frequency and pressure in the region below the white dashed curve. The maximum rotation frequency is $\omega_r = 2\pi \times 20$ MHz in this experiment, which is limited by the $\pi$-phase shifter (Mini-Circuits, ZFSCJ-2-2-S) used to generate the signals on the four electrodes. Under favorable conditions, the rotational motion can achieve a frequency exceeding 10 GHz at the pressure of $10^{-6}$ Torr based on the dashed line in Supplementary Fig. \ref{fig:s5}(e).

\section{Berry phase of rotating NV electron spins}

\begin{figure*}
	\includegraphics[width=\textwidth]{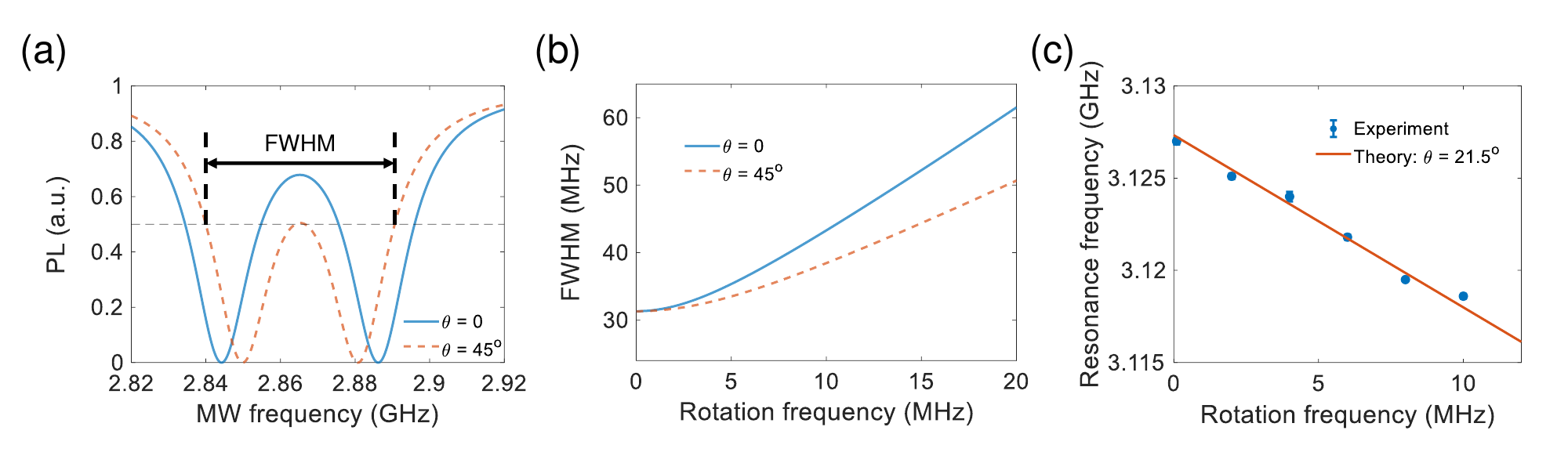}
	\caption{\label{fig:s6} \rw{Berry phase induced} by a rotating nanodiamond. (a) Theoretically calculated ODMR of NV centers with different orientations at the rotation frequency of 20 MHz. The blue solid, red dashed and orange dash-dotted curves are calculated at $\theta = 0$ and $\theta = 45^\circ$, respectively. (b) Theoretically calculated FWHM of the ODMR as a function of rotation frequency. The blue solid curve \rw{and the} red dashed curve are calculated for $\theta = 0$ and $\theta = 45^\circ$, respectively. \rw{(c) Experimental results of the frequency shift due to the Berry phase induced by counterclockwise rotation (blue circles), and theoretical calculated resonance frequency as a function of the rotation frequency at $\theta = 21.5^\circ$ (red curve).}}
\end{figure*}

\subsection{Without an external magnetic field}
\rw{In a rotating diamond, the embedded NV centers follow the rotation of the particle with an angular frequency of $\omega_r$. Considering an arbitrary NV center in a diamond at the time of $t$, the angle between the NV axis and $z$ axis is $\theta$, and the azimuth angle is $\phi(t) = \omega_r t$ relative to $x$ axis. In the absence of an external magnetic field and neglecting strain effects, the Hamiltonian of the rotating NV center in the laboratory frame can be written as \cite{Maclaurin2012}
	\begin{equation}
		\begin{array}{l}
			{H_{0,lab}} = \frac{1}{\hbar }R\left( t \right)DS_z^2{R^\dag }\left( t \right) = \frac{1}{\hbar }{e^{ - i\phi {S_z}}}{e^{ - i\theta {S_y}}}DS_z^2{e^{i\theta {S_y}}}{e^{i\phi {S_z}}}\\
			= D\hbar \left( {\begin{array}{*{20}{c}}
					{{{\cos }^2}\theta  + \frac{{{{\sin }^2}\theta }}{2}}&{\frac{{{e^{ - i\phi }}\cos \theta \sin \theta }}{{\sqrt 2 }}}&{\frac{{{e^{ - 2i\phi }}{{\sin }^2}\theta }}{2}}\\
					{\frac{{{e^{i\phi }}\cos \theta \sin \theta }}{{\sqrt 2 }}}&{{{\sin }^2}\theta }&{ - \frac{{{e^{ - i\phi }}\cos \theta \sin \theta }}{{\sqrt 2 }}}\\
					{\frac{{{e^{2i\phi }}{{\sin }^2}\theta }}{2}}&{ - \frac{{{e^{i\phi }}\cos \theta \sin \theta }}{{\sqrt 2 }}}&{{{\cos }^2}\theta  + \frac{{{{\sin }^2}\theta }}{2}}
			\end{array}} \right)
		\end{array}
		\label{eq:h0lab},
	\end{equation}
	where $D$ is the zero-field splitting, $R\left( t \right) = {R_z}\left( {\phi \left( t \right)} \right){R_y}\left( \theta  \right)$ is the rotation transformation, and ${R_y}\left( \theta  \right) = \exp \left( { - i\theta {S_y}} \right)$ (${R_z}\left( \phi  \right) = \exp \left( { - i\phi {S_z}} \right)$) expresses the rotation of spin around the y (z) axis in terms of $\theta$ ($\phi$), ${\bf{S}}$ is the spin operator. The Hamiltonian possesses three eigenstates ${\left| {{m_s},t} \right\rangle _{lab}} = R\left( t \right){\left| {{m_s},0} \right\rangle _{lab}}$ ($m_s = 0, \pm 1$),
	\begin{equation}
		\begin{array}{l}
			{\left| {1,t} \right\rangle _{lab}} = R\left( t \right)\left( {\begin{array}{*{20}{c}}
					1\\
					0\\
					0
			\end{array}} \right) = \left( {\begin{array}{*{20}{c}}
					{{e^{ - i\phi }}{{\cos }^2}\frac{\theta }{2}}\\
					{\frac{{\sin \theta }}{{\sqrt 2 }}}\\
					{{e^{i\phi }}{{\sin }^2}\frac{\theta }{2}}
			\end{array}} \right)\\
			{\left| {0,t} \right\rangle _{lab}} = R\left( t \right)\left( {\begin{array}{*{20}{c}}
					0\\
					1\\
					0
			\end{array}} \right) = \left( {\begin{array}{*{20}{c}}
					{ - \frac{{{e^{ - i\phi }}\sin \theta }}{{\sqrt 2 }}}\\
					{\cos \theta }\\
					{\frac{{{e^{i\phi }}\sin \theta }}{{\sqrt 2 }}}
			\end{array}} \right)\\
			{\left| { - 1,t} \right\rangle _{lab}} = R\left( t \right)\left( {\begin{array}{*{20}{c}}
					0\\
					0\\
					1
			\end{array}} \right) = \left( {\begin{array}{*{20}{c}}
					{{e^{ - i\phi }}{{\sin }^2}\frac{\theta }{2}}\\
					{ - \frac{{\sin \theta }}{{\sqrt 2 }}}\\
					{{e^{i\phi }}{{\cos }^2}\frac{\theta }{2}}
			\end{array}} \right)
		\end{array}
		\label{eq:h0evlab}.
\end{equation}}

\rw{For a quantum system in an eigenstate,  the system remains in the eigenstate and acquires a phase factor during an adiabatic evolution of the Hamiltonian. This factor arises from both the state's time evolution and the variation of the eigenstate with the changing Hamiltonian. The second term specifically corresponds to the Berry phase. Hence, the expression for the time-dependent spin state is \cite{Chudo2021}
	\begin{equation}
		{e^{i{\gamma _{{m_s}}}}}{e^{ - i{H_{0,lab}}t/\hbar }}{\left| {{m_s},t} \right\rangle _{lab}} = {e^{i{\gamma _{{m_s}}}}}{e^{ - i{H_{0,lab}}t/\hbar }}{e^{ - i\phi {S_z}}}{e^{ - i\theta {S_y}}}{\left| {{m_s},0} \right\rangle _{lab}}
		\label{eq:h0sslab},
	\end{equation}
	where ${\gamma _{{m_s}}}$ is the Berry phase. Here, the diamond particle rotates around the $z$ axis with a constant $\theta$, the Berry phase can be calculated as \cite{Chudo2021}
	\begin{equation}
		{\gamma _{{m_s}}} = i\int_0^t {_{lab}\left\langle {{m_s},t'} \right|\frac{\partial }{{\partial t'}}{{\left| {{m_s},t'} \right\rangle }_{lab}}dt'}  = {m_s}{\omega _r}t\cos \theta 
		\label{eq:berryphase}.
	\end{equation}
	The Berry phase of Eq.~\ref{eq:berryphase} is calculated for an open-path, which is gauge-dependent. However, for a closed loop, the Berry phase is gauge-invariant and can be expressed as ${m_s}\left[ { - 2\pi \left( {1 - \cos \theta } \right)} \right]$. The result is equivalent to Eq.~\ref{eq:berryphase} of ${m_s} \left(2\pi \cos \theta \right)$.}

\rw{The spin state of the NV center is observed through the interaction with a microwave magnetic field. In our experiment, the direction of the microwave is in the $yz$-plane and forms a slight angle $\theta'$ relative to the $z$ axis, resulting from the asymmetric design of the waveguide. The Hamiltonian of the microwave in the laboratory frame can be written as
	\begin{equation}
		{H_{MW,lab}} = g{\mu _B}{B_{MW}}\cos \left( {{\omega _{MW}}t} \right)\left( {{S_z}\cos \theta ' + {S_y}\sin \theta '} \right) = {H_{MW,z,lab}} + {H_{MW,y,lab}}
		\label{eq:microwavelab},
	\end{equation}
	which contains two components, the longitudinal term ${H_{MW,z,lab}} = g{\mu _B}{B_{MW}}\cos \left( {{\omega _{MW}}t} \right){S_z}\cos \theta '$ and the transverse term ${H_{MW,y,lab}} = g{\mu _B}{B_{MW}}\cos \left( {{\omega _{MW}}t} \right){S_y}\sin \theta '$. First, we consider the longitudinal term ${H_{MW,z,lab}}$. The expected value of the spin states can be expressed as
	\begin{equation}
		\begin{array}{l}
			_{lab}\left\langle { \pm 1,t} \right|{e^{i{H_{0,lab}}t/\hbar }}{e^{ - i{\gamma _{ \pm 1}}}}{H_{MW,z,lab}}{e^{i{\gamma _0}}}{e^{ - i{H_{0,lab}}t/\hbar }}{\left| {0,t} \right\rangle _{lab}}\\
			= g{\mu _B}{B_{MW}}\cos \left( {{\omega _{MW}}t} \right)\cos {{\theta '}_{lab}}\left\langle { \pm 1,0} \right|{e^{i\theta {S_y}}}{e^{i\phi {S_z}}}{e^{i{H_{0,lab}}t/\hbar }}{e^{ - i{\gamma _{ \pm 1}}}}{S_z}{e^{i{\gamma _0}}}{e^{ - i{H_{0,lab}}t/\hbar }}{e^{ - i\phi {S_z}}}{e^{ - i\theta {S_y}}}{\left| {0,0} \right\rangle _{lab}}\\
			= g{\mu _B}{B_{MW}}\cos \left( {{\omega _{MW}}t} \right)\cos \theta '{e^{ - i\left( {{\gamma _{ \pm 1}} - {\gamma _0}} \right)}}{e^{i\left( {{E_{ \pm 1}} - {E_0}} \right)t/\hbar }}_{lab}\left\langle { \pm 1,0} \right|{e^{i\theta {S_y}}}{e^{i\phi {S_z}}}{S_z}{e^{ - i\phi {S_z}}}{e^{ - i\theta {S_y}}}{\left| {0,0} \right\rangle _{lab}}\\
			= \frac{1}{2}g{\mu _B}{B_{MW}}\cos \theta '\left( {{e^{i{\omega _{MW}}t}} + {e^{ - i{\omega _{MW}}t}}} \right){e^{ \mp i{\omega _r}t\cos \theta }}{e^{iDt}}_{lab}\left\langle { \pm 1,0} \right|{e^{i\theta {S_y}}}{S_z}{e^{ - i\theta {S_y}}}{\left| {0,0} \right\rangle _{lab}}\\
			= \frac{1}{2}g{\mu _B}{B_{MW}}\cos \theta '{\left[ {{e^{i\left( {{\omega _{MW}} \mp {\omega _r}\cos \theta  + D} \right)t}} + {e^{i\left( { - {\omega _{MW}} \mp {\omega _r}\cos \theta  + D} \right)t}}} \right]_{lab}}\left\langle { \pm 1,0} \right|{e^{i\theta {S_y}}}{S_z}{e^{ - i\theta {S_y}}}{\left| {0,0} \right\rangle _{lab}}\\
			\approx \frac{1}{2}g{\mu _B}{B_{MW}}\cos \theta '{e^{i\left( { - {\omega _{MW}} + D \mp {\omega _r}\cos \theta } \right)t}}_{lab}\left\langle { \pm 1,0} \right|{e^{i\theta {S_y}}}{S_z}{e^{ - i\theta {S_y}}}{\left| {0,0} \right\rangle _{lab}}
		\end{array}
		\label{eq:h0mwzlab},
	\end{equation}
	where the $E_{m_s}$ is the corresponding eigenvalue of the Hamiltonian $H_{0,lab}$ for the spin state $\left| {m_s,t} \right\rangle$. According to Eq.~\ref{eq:h0mwzlab}, the transformation of spin states from ${\left| {{m_s} = 0} \right\rangle _{lab}}$ to ${\left| {{m_s} =  \pm 1} \right\rangle _{lab}}$ can be driven by a microwave operating at the resonance frequency of $D \mp {\omega _r}\cos \theta$, where the frequency shift $\mp {\omega _r}\cos \theta$ is attributed to the Berry phase.}

\rw{Regarding the second part of the Hamiltonian in the transverse direction, $H_{MW,y,lab}$, the interaction between the microwave and the spin states can be formulated as
	\begin{equation}
		\begin{array}{l}
			_{lab}\left\langle { \pm 1,t} \right|{e^{i{H_{0,lab}}t/\hbar }}{e^{ - i{\gamma _{ \pm 1}}}}{H_{MW,y,lab}}{e^{i{\gamma _0}}}{e^{ - i{H_{0,lab}}t/\hbar }}{\left| {0,t} \right\rangle _{lab}}\\
			= g{\mu _B}{B_{MW}}\cos \left( {{\omega _{MW}}t} \right)\sin {{\theta '}_{lab}}\left\langle { \pm 1,0} \right|{e^{i\theta {S_y}}}{e^{i\phi {S_z}}}{e^{i{H_{0,lab}}t/\hbar }}{e^{ - i{\gamma _{ \pm 1}}}}{S_y}{e^{i{\gamma _0}}}{e^{ - i{H_{0,lab}}t/\hbar }}{e^{ - i\phi {S_z}}}{e^{ - i\theta {S_y}}}{\left| {0,0} \right\rangle _{lab}}\\
			= g{\mu _B}{B_{MW}}\cos \left( {{\omega _{MW}}t} \right)\sin \theta '{e^{ - i\left( {{\gamma _{ \pm 1}} - {\gamma _0}} \right)}}{e^{i\left( {{E_{ \pm 1}} - {E_0}} \right)t/\hbar }}_{lab}\left\langle { \pm 1,0} \right|{e^{i\theta {S_y}}}{e^{i\phi {S_z}}}{S_y}{e^{ - i\phi {S_z}}}{e^{ - i\theta {S_y}}}{\left| {0,0} \right\rangle _{lab}}\\
			= \frac{1}{2}g{\mu _B}{B_{MW}}\sin \theta '\left( {{e^{i{\omega _{MW}}t}} + {e^{ - i{\omega _{MW}}t}}} \right){e^{ \mp i{\omega _r}t\cos \theta }}{e^{iDt}}\\
			{ \times _{lab}}\left\langle { \pm 1,0} \right|{e^{i\theta {S_y}}}\frac{1}{{2i}}\left( {{S_ + }{e^{i{\omega _r}t}} - {S_ - }{e^{ - i{\omega _r}t}}} \right){e^{ - i\theta {S_y}}}{\left| {0,0} \right\rangle _{lab}}
		\end{array}
		\label{eq:h0mwylab}.
	\end{equation}
	The expected values are written as
	\begin{equation}
		\begin{array}{l}
			_{lab}\left\langle { + 1,t} \right|{e^{i{H_{0,lab}}t/\hbar }}{e^{ - i{\gamma _{ + 1}}}}{H_{MW,y,lab}}{e^{i{\gamma _0}}}{e^{ - i{H_{0,lab}}t/\hbar }}{\left| {0,t} \right\rangle _{lab}}\\
			= \frac{1}{{4i}}g{\mu _B}{B_{MW}}\sin \theta '\left[ {{e^{i\left( {{\omega _{MW}} - {\omega _r}\cos \theta  + D + {\omega _r}} \right)t}} + {e^{i\left( { - {\omega _{MW}} - {\omega _r}\cos \theta  + D + {\omega _r}} \right)t}}} \right]\\
			{ \times _{lab}}\left\langle { + 1,0} \right|{e^{i\theta {S_y}}}{S_ + }{e^{ - i\theta {S_y}}}{\left| {0,0} \right\rangle _{lab}}\\
			\approx \frac{1}{{4i}}g{\mu _B}{B_{MW}}\sin \theta '{e^{i\left( { - {\omega _{MW}} + D + {\omega _r} - {\omega _r}\cos \theta } \right)t}}_{lab}\left\langle { + 1,0} \right|{e^{i\theta {S_y}}}{S_ + }{e^{ - i\theta {S_y}}}{\left| {0,0} \right\rangle _{lab}}
		\end{array}
		\label{eq:h0mwylab1},
	\end{equation}
	\begin{equation}
		\begin{array}{l}
			_{lab}\left\langle { - 1,t} \right|{e^{i{H_{0,lab}}t/\hbar }}{e^{ - i{\gamma _{ - 1}}}}{H_{MW,y,lab}}{e^{i{\gamma _0}}}{e^{ - i{H_{0,lab}}t/\hbar }}{\left| {0,t} \right\rangle _{lab}}\\
			=  - \frac{1}{{4i}}g{\mu _B}{B_{MW}}\sin \theta '\left[ {{e^{i\left( {{\omega _{MW}} + {\omega _r}\cos \theta  + D - {\omega _r}} \right)t}} + {e^{i\left( { - {\omega _{MW}} + {\omega _r}\cos \theta  + D - {\omega _r}} \right)t}}} \right]\\
			{ \times _{lab}}\left\langle { - 1,0} \right|{e^{i\theta {S_y}}}{S_ - }{e^{ - i\theta {S_y}}}{\left| {0,0} \right\rangle _{lab}}\\
			\approx  - \frac{1}{{4i}}g{\mu _B}{B_{MW}}\sin \theta '{e^{i\left( { - {\omega _{MW}} + D - {\omega _r} + {\omega _r}\cos \theta } \right)t}}_{lab}\left\langle { - 1,0} \right|{e^{i\theta {S_y}}}{S_ - }{e^{ - i\theta {S_y}}}{\left| {0,0} \right\rangle _{lab}}
		\end{array}
		\label{eq:h0mwylab2}.
	\end{equation}
	Utilizing Eq.~\ref{eq:h0mwylab1} and Eq.~\ref{eq:h0mwylab2}, the resonance frequency of microwave for transforming the spin state from ${\left| {{m_s} = 0} \right\rangle _{lab}}$ to ${\left| {{m_s} =  \pm 1} \right\rangle _{lab}}$ is $D \pm {\omega _r}\left( {1 - \cos \theta } \right)$. In addition to the frequency shift of $\mp {\omega _r}\cos \theta$ caused by the Berry phase, there is another term of ${\omega _r}$ coming from the rotational Doppler effect \cite{Chudo2021}. In our experiment, the angle $\theta '$ of the microwave relate to the $z$ axis is approximately 8.5$^\circ$. Consequently, the dominant transition probability arises from the longitudinal component, characterized by a frequency shift of $\mp {\omega _r}\cos \theta$ due to the Berry phase.}

The energy levels of NV centers with four orientations are degenerate in the absence of an external magnetic field. \rw{When the nanodiamond undergoes rotation, the electron spin resonance frequency experiences a shift due to the Berry phase, and this shift depends on the angle $\theta$ between the NV axis and the rotation axis.} The electron spin resonance frequencies of NV spins along different orientations become non-degenerate. The ODMR of NV at different orientations are theoretically calculated by Eq.~\ref{eq:h0mwzlab} at the rotation frequency of 20 MHz (Supplementary Fig. \ref{fig:s6}(a)). The orientations corresponding to the blue solid curve \rw{and the} red dashed curve are $\theta = 0^\circ$ and $\theta = 45^\circ$, respectively. The intrinsic linewidth is $2\pi \times 19$ MHz, and the strain effect splitting $E$ is $2\pi \times 6.7$ MHz. The eight dips are not separated in the ODMR spectrum at a rotation frequency of a few MHz because of the large linewidth. Here we use the FWHM parameter of the ODMR spectrum to indicate the \rw{frequency shift by the Berry phase of a rotating NV center}. The FWHM of the ODMR is mainly determined by the splitting of the NV centers that have the smallest $\theta$. Supplementary Fig. \ref{fig:s6}(b) shows the FWHM of the ODMR as a function of rotation frequency. The NV centers, which have the smallest $\theta$, show the highest sensitivity of the \rw{frequency shift due to the Berry phase }.

\subsection{With an external magnetic field}
\rw{To precisely measure the frequency shift induced by the Berry phase of a rotating NV center, an external magnetic field $\bf{B}$ along the z direction can be applied. This serves to distinguish the energy levels of NV centers in four different orientations. The Hamiltonian of a NV center in the laboratory frame with the external magnetic field can be expressed as
	\begin{equation}
		\begin{array}{l}
			{H_{lab}} = {H_{0,lab}} + g{\mu _B}B{S_z} = \frac{1}{\hbar }{e^{ - i\phi {S_z}}}{e^{ - i\theta {S_y}}}DS_z^2{e^{i\theta {S_y}}}{e^{i\phi {S_z}}} + g{\mu _B}B{S_z}\\
			= D\hbar \left( {\begin{array}{*{20}{c}}
					{{{\cos }^2}\theta  + \frac{{{{\sin }^2}\theta }}{2} + \frac{{g{\mu _B}B}}{D}}&{\frac{{{e^{ - i\phi }}\cos \theta \sin \theta }}{{\sqrt 2 }}}&{\frac{{{e^{ - 2i\phi }}{{\sin }^2}\theta }}{2}}\\
					{\frac{{{e^{i\phi }}\cos \theta \sin \theta }}{{\sqrt 2 }}}&{{{\sin }^2}\theta }&{ - \frac{{{e^{ - i\phi }}\cos \theta \sin \theta }}{{\sqrt 2 }}}\\
					{\frac{{{e^{2i\phi }}{{\sin }^2}\theta }}{2}}&{ - \frac{{{e^{i\phi }}\cos \theta \sin \theta }}{{\sqrt 2 }}}&{{{\cos }^2}\theta  + \frac{{{{\sin }^2}\theta }}{2} - \frac{{g{\mu _B}B}}{D}}
			\end{array}} \right)
		\end{array}
		\label{eq:hlab}.
	\end{equation}
	In the rotating frame, the Hamiltonian of the NV center can be calculated by a unitary transformation,
	\begin{equation}
		\begin{array}{l}
			{H_{rot}} = U{H_{lab}}{U^\dag } + i{\partial _t}U{U^\dag } = {e^{i\theta {S_y}}}{e^{i\phi {S_z}}}{H_{lab}}{e^{ - i\phi {S_z}}}{e^{ - i\theta {S_y}}} + i{\partial _t}{e^{i\theta {S_y}}}{e^{i\phi {S_z}}}{e^{ - i\phi {S_z}}}{e^{ - i\theta {S_y}}}\\
			= \hbar \left( {\begin{array}{*{20}{c}}
					{D + g{\mu _B}B\cos \theta }&{ - \frac{{g{\mu _B}B\sin \theta }}{{\sqrt 2 }}}&0\\
					{ - \frac{{g{\mu _B}B\sin \theta }}{{\sqrt 2 }}}&0&{ - \frac{{g{\mu _B}B\sin \theta }}{{\sqrt 2 }}}\\
					0&{ - \frac{{g{\mu _B}B\sin \theta }}{{\sqrt 2 }}}&{D - g{\mu _B}B\cos \theta }
			\end{array}} \right) + \hbar \left( {\begin{array}{*{20}{c}}
					{ - {\omega _r}\cos \theta }&{\frac{{{\omega _r}\sin \theta }}{{\sqrt 2 }}}&0\\
					{\frac{{{\omega _r}\sin \theta }}{{\sqrt 2 }}}&0&{\frac{{{\omega _r}\sin \theta }}{{\sqrt 2 }}}\\
					0&{\frac{{{\omega _r}\sin \theta }}{{\sqrt 2 }}}&{{\omega _r}\cos \theta }
			\end{array}} \right)
		\end{array}
		\label{eq:hrot},
	\end{equation}
	where the unitary operator is defined as $U = {e^{i\theta {S_y}}}{e^{i\phi {S_z}}}$. The second term on the right side of Eq.~\ref{eq:hrot} represents Zeeman interaction arising from the pseudo-magnetic field due to the rotation of the NV center. In the case of an adiabatic process, ${\omega _r} \ll D - g{\mu _B}B\cos \theta $, the second term is significantly weaker than the first term, and can be treated as a perturbation. We neglect the off-diagonal terms in the first component since $g{\mu _B}B \ll D$, which are too small to induce significant mixing of the NV spin states.
	\begin{equation}
		{H_{rot}} \approx \hbar \left( {\begin{array}{*{20}{c}}
				{D + g{\mu _B}B\cos \theta }&0&0\\
				0&0&0\\
				0&0&{D - g{\mu _B}B\cos \theta }
		\end{array}} \right) + \hbar \left( {\begin{array}{*{20}{c}}
				{ - {\omega _r}\cos \theta }&{\frac{{{\omega _r}\sin \theta }}{{\sqrt 2 }}}&0\\
				{\frac{{{\omega _r}\sin \theta }}{{\sqrt 2 }}}&0&{\frac{{{\omega _r}\sin \theta }}{{\sqrt 2 }}}\\
				0&{\frac{{{\omega _r}\sin \theta }}{{\sqrt 2 }}}&{{\omega _r}\cos \theta }
		\end{array}} \right)
		\label{eq:hrot1}.
	\end{equation}
	Therefore, the Hamiltonian in the rotating frame possesses three eigenstates, ${\left| {{m_s}} \right\rangle _{rot}}$ ($m_s = 0, \pm 1$):
	\begin{equation}
		\begin{array}{l}
			{\left| { + 1} \right\rangle _{rot}} = \left( {\begin{array}{*{20}{c}}
					1\\
					0\\
					0
			\end{array}} \right)\\
			{\left| 0 \right\rangle _{rot}} = \left( {\begin{array}{*{20}{c}}
					0\\
					1\\
					0
			\end{array}} \right)\\
			{\left| { - 1} \right\rangle _{rot}} = \left( {\begin{array}{*{20}{c}}
					0\\
					0\\
					1
			\end{array}} \right)
		\end{array}
		\label{eq:hevrot},
	\end{equation}
	and the corresponding eigenvalues are $\hbar \left( {D + g{\mu _B}B\cos \theta } \right)$, 0, $\hbar \left( {D + g{\mu _B}B\cos \theta } \right)$, respectively. The new Hamiltonian in the laboratory frame can be transformed by applying the rotation transformation $R\left( t \right) = {e^{ - i\phi {S_z}}}{e^{ - i\theta {S_y}}}$,
	\begin{equation}
		\begin{array}{l}
			{H_{lab}} = R{H_{rot}}{R^\dag } + i{\partial _t}R{R^\dag } = {e^{ - i\phi {S_z}}}{e^{ - i\theta {S_y}}}{H_{rot}}{e^{i\theta {S_y}}}{e^{i\phi {S_z}}} + i{\partial _t}{e^{ - i\phi {S_z}}}{e^{ - i\theta {S_y}}}{e^{i\theta {S_y}}}{e^{i\phi {S_z}}}\\
			= \hbar \left( {\begin{array}{*{20}{c}}
					{D\frac{{1 + {{\cos }^2}\theta }}{2} + g{\mu _B}B{{\cos }^2}\theta }&{\frac{{{e^{ - i\phi }}\left( {D + g{\mu _B}B} \right)\cos \theta \sin \theta }}{{\sqrt 2 }}}&{\frac{{{e^{ - 2i\phi }}D{{\sin }^2}\theta }}{2}}\\
					{\frac{{{e^{i\phi }}\left( {D + g{\mu _B}B} \right)\cos \theta \sin \theta }}{{\sqrt 2 }}}&{D{{\sin }^2}\theta }&{ - \frac{{{e^{ - i\phi }}\left( {D - g{\mu _B}B} \right)\cos \theta \sin \theta }}{{\sqrt 2 }}}\\
					{\frac{{{e^{2i\phi }}D{{\sin }^2}\theta }}{2}}&{ - \frac{{{e^{i\phi }}\left( {D - g{\mu _B}B} \right)\cos \theta \sin \theta }}{{\sqrt 2 }}}&{D\frac{{1 + {{\cos }^2}\theta }}{2} - g{\mu _B}B{{\cos }^2}\theta }
			\end{array}} \right)
		\end{array}
		\label{eq:hlab1}.
	\end{equation}
	The eigenstates of the Hamiltonian in the laboratory frame also can be calculated through the rotation transformation,
	\begin{equation}
		\begin{array}{l}
			{\left| { + 1,t} \right\rangle _{lab}} = R\left( t \right){\left| { + 1} \right\rangle _{rot}} = \left( {\begin{array}{*{20}{c}}
					{{e^{ - i\phi }}{{\cos }^2}\frac{\theta }{2}}\\
					{\frac{{\sin \theta }}{{\sqrt 2 }}}\\
					{{e^{i\phi }}{{\sin }^2}\frac{\theta }{2}}
			\end{array}} \right)\\
			{\left| {0,t} \right\rangle _{lab}} = R\left( t \right){\left| 0 \right\rangle _{rot}} = \left( {\begin{array}{*{20}{c}}
					{ - \frac{{{e^{ - i\phi }}\sin \theta }}{{\sqrt 2 }}}\\
					{\cos \theta }\\
					{\frac{{{e^{i\phi }}\sin \theta }}{{\sqrt 2 }}}
			\end{array}} \right)\\
			{\left| { - 1,t} \right\rangle _{lab}} = R\left( t \right){\left| { - 1} \right\rangle _{rot}} = \left( {\begin{array}{*{20}{c}}
					{{e^{ - i\phi }}{{\sin }^2}\frac{\theta }{2}}\\
					{ - \frac{{\sin \theta }}{{\sqrt 2 }}}\\
					{{e^{i\phi }}{{\cos }^2}\frac{\theta }{2}}
			\end{array}} \right)
		\end{array}
		\label{eq:hevlab},
	\end{equation}
	which are same as the eigenstates of the Hamiltonian without the external magnetic field (Eq.~\ref{eq:h0evlab}). Thus, the Berry phase of the rotating NV center is:
	\begin{equation}
		{\gamma _{{m_s}}} = {m_s}{\omega _r}t\cos \theta
		\label{eq:berryphase1}
	\end{equation}.}

\rw{Similarly, the interaction between microwave and the time-dependent spin states is still divided into two components. For the longitudinal component, it can be expressed as
	\begin{equation}
		\begin{array}{l}
			_{lab}\left\langle { \pm 1,t} \right|{e^{i{H_{lab}}t/\hbar }}{e^{ - i{\gamma _{ \pm 1}}}}{H_{MW,z,lab}}{e^{i{\gamma _0}}}{e^{ - i{H_{lab}}t/\hbar }}{\left| {0,t} \right\rangle _{lab}}\\
			= g{\mu _B}{B_{MW}}\cos \left( {{\omega _{MW}}t} \right)\cos {{\theta '}_{lab}}\left\langle { \pm 1,0} \right|{e^{i\theta {S_y}}}{e^{i\phi {S_z}}}{e^{i{H_{lab}}t/\hbar }}{e^{ - i{\gamma _{ \pm 1}}}}{S_z}{e^{i{\gamma _0}}}{e^{ - i{H_{lab}}t/\hbar }}{e^{ - i\phi {S_z}}}{e^{ - i\theta {S_y}}}{\left| {0,0} \right\rangle _{lab}}\\
			= g{\mu _B}{B_{MW}}\cos \left( {{\omega _{MW}}t} \right)\cos \theta '{e^{ - i\left( {{\gamma _{ \pm 1}} - {\gamma _0}} \right)}}{e^{i\left( {{E_{B, \pm 1}} - {E_{B,0}}} \right)t/\hbar }}_{lab}\left\langle { \pm 1,0} \right|{e^{i\theta {S_y}}}{e^{i\phi {S_z}}}{S_z}{e^{ - i\phi {S_z}}}{e^{ - i\theta {S_y}}}{\left| {0,0} \right\rangle _{lab}}\\
			= \frac{1}{2}g{\mu _B}{B_{MW}}\cos \theta '\left( {{e^{i{\omega _{MW}}t}} + {e^{ - i{\omega _{MW}}t}}} \right){e^{ \mp i{\omega _r}t\cos \theta }}{e^{i\left( {D \pm g{\mu _B}B\cos \theta } \right)t}}_{lab}\left\langle { \pm 1,0} \right|{e^{i\theta {S_y}}}{S_z}{e^{ - i\theta {S_y}}}{\left| {0,0} \right\rangle _{lab}}\\
			= \frac{1}{2}g{\mu _B}{B_{MW}}\cos \theta '\left[ {{e^{i\left( {{\omega _{MW}} \mp {\omega _r}\cos \theta  + D \pm g{\mu _B}B\cos \theta } \right)t}} + {e^{i\left( { - {\omega _{MW}} \mp {\omega _r}\cos \theta  + D \pm g{\mu _B}B\cos \theta } \right)t}}} \right]\\
			{ \times _{lab}}\left\langle { \pm 1,0} \right|{e^{i\theta {S_y}}}{S_z}{e^{ - i\theta {S_y}}}{\left| {0,0} \right\rangle _{lab}}\\
			\approx \frac{1}{2}g{\mu _B}{B_{MW}}\cos \theta '{e^{i\left( { - {\omega _{MW}} + D \pm g{\mu _B}B\cos \theta  \mp {\omega _r}\cos \theta } \right)t}}_{lab}\left\langle { \pm 1,0} \right|{e^{i\theta {S_y}}}{S_z}{e^{ - i\theta {S_y}}}{\left| {0,0} \right\rangle _{lab}}
		\end{array}
		\label{hmwzlab},
	\end{equation}
	where the $E_{B,m_s}$ is the eigenvalue of the Hamiltonian $H_{lab}$ for the spin state ${\left| {{m_s},t} \right\rangle _{lab}}$. The spin resonance frequency, transformed from ${\left| {{m_s} = 0} \right\rangle _{lab}}$ to ${\left| {{m_s} =  \pm 1} \right\rangle _{lab}}$ is $D \pm g{\mu _B}B\cos \theta  \mp {\omega _r}\cos \theta $. The frequency shift due to the Berry phase is $\mp {\omega _r}\cos \theta$.}

\rw{The transverse component can be expressed as
	\begin{equation}
		\begin{array}{l}
			_{lab}\left\langle { \pm 1,t} \right|{e^{i{H_{lab}}t/\hbar }}{e^{ - i{\gamma _{ \pm 1}}}}{H_{MW,y,lab}}{e^{i{\gamma _0}}}{e^{ - i{H_{lab}}t/\hbar }}{\left| {0,t} \right\rangle _{lab}}\\
			= g{\mu _B}{B_{MW}}\cos \left( {{\omega _{MW}}t} \right)\sin {{\theta '}_{lab}}\left\langle { \pm 1,0} \right|{e^{i\theta {S_y}}}{e^{i\phi {S_z}}}{e^{i{H_{lab}}t/\hbar }}{e^{ - i{\gamma _{ \pm 1}}}}{S_y}{e^{i{\gamma _0}}}{e^{ - i{H_{lab}}t/\hbar }}{e^{ - i\phi {S_z}}}{e^{ - i\theta {S_y}}}{\left| {0,0} \right\rangle _{lab}}\\
			= g{\mu _B}{B_{MW}}\cos \left( {{\omega _{MW}}t} \right)\sin \theta '{e^{ - i\left( {{\gamma _{ \pm 1}} - {\gamma _0}} \right)}}{e^{i\left( {{E_{B, \pm 1}} - {E_{B,0}}} \right)t/\hbar }}_{lab}\left\langle { \pm 1,0} \right|{e^{i\theta {S_y}}}{e^{i\phi {S_z}}}{S_y}{e^{ - i\phi {S_z}}}{e^{ - i\theta {S_y}}}{\left| {0,0} \right\rangle _{lab}}\\
			= \frac{1}{2}g{\mu _B}{B_{MW}}\sin \theta '\left( {{e^{i{\omega _{MW}}t}} + {e^{ - i{\omega _{MW}}t}}} \right){e^{ \mp i{\omega _r}t\cos \theta }}{e^{i\left( {D \pm g{\mu _B}B\cos \theta } \right)t}}\\
			{ \times _{lab}}\left\langle { \pm 1,0} \right|{e^{i\theta {S_y}}}\frac{1}{{2i}}\left( {{S_ + }{e^{i{\omega _r}t}} - {S_ - }{e^{ - i{\omega _r}t}}} \right){e^{ - i\theta {S_y}}}{\left| {0,0} \right\rangle _{lab}}
		\end{array}
		\label{eq:hmwylab}.
	\end{equation}
	The expected values are written as
	\begin{equation}
		\begin{array}{l}
			_{lab}\left\langle { + 1,t} \right|{e^{i{H_{lab}}t/\hbar }}{e^{ - i{\gamma _{ + 1}}}}{H_{MW,y,lab}}{e^{i{\gamma _0}}}{e^{ - i{H_{lab}}t/\hbar }}{\left| {0,t} \right\rangle _{lab}}\\
			= \frac{1}{{4i}}g{\mu _B}{B_{MW}}\sin \theta '\left[ {{e^{i\left( {{\omega _{MW}} - {\omega _r}\cos \theta  + D + g{\mu _B}B\cos \theta  + {\omega _r}} \right)t}} + {e^{i\left( { - {\omega _{MW}} - {\omega _r}\cos \theta  + D + g{\mu _B}B\cos \theta  + {\omega _r}} \right)t}}} \right]\\
			{ \times _{lab}}\left\langle { + 1,0} \right|{e^{i\theta {S_y}}}{S_ + }{e^{ - i\theta {S_y}}}{\left| {0,0} \right\rangle _{lab}}\\
			\approx \frac{1}{{4i}}g{\mu _B}{B_{MW}}\sin \theta '{e^{i\left( { - {\omega _{MW}} - {\omega _r}\cos \theta  + D + g{\mu _B}B\cos \theta  + {\omega _r}} \right)t}}_{lab}\left\langle { + 1,0} \right|{e^{i\theta {S_y}}}{S_ + }{e^{ - i\theta {S_y}}}{\left| {0,0} \right\rangle _{lab}}
		\end{array}
		\label{eq:hmwylab1},
	\end{equation}
	\begin{equation}
		\begin{array}{l}
			_{lab}\left\langle { - 1,t} \right|{e^{i{H_{lab}}t/\hbar }}{e^{ - i{\gamma _{ - 1}}}}{H_{MW,y,lab}}{e^{i{\gamma _0}}}{e^{ - i{H_{lab}}t/\hbar }}{\left| {0,t} \right\rangle _{lab}}\\
			=  - \frac{1}{{4i}}g{\mu _B}{B_{MW}}\sin \theta '\left[ {{e^{i\left( {{\omega _{MW}} + {\omega _r}\cos \theta  + D - g{\mu _B}B\cos \theta  - {\omega _r}} \right)t}} + {e^{i\left( { - {\omega _{MW}} + {\omega _r}\cos \theta  + D - g{\mu _B}B\cos \theta  - {\omega _r}} \right)t}}} \right]\\
			{ \times _{lab}}\left\langle { - 1,0} \right|{e^{i\theta {S_y}}}{S_ - }{e^{ - i\theta {S_y}}}{\left| {0,0} \right\rangle _{lab}}\\
			\approx  - \frac{1}{{4i}}g{\mu _B}{B_{MW}}\sin \theta '{e^{i\left( { - {\omega _{MW}} + {\omega _r}\cos \theta  + D - g{\mu _B}B\cos \theta  - {\omega _r}} \right)t}}_{lab}\left\langle { - 1,0} \right|{e^{i\theta {S_y}}}{S_ - }{e^{ - i\theta {S_y}}}{\left| {0,0} \right\rangle _{lab}}
		\end{array}
		\label{eq:hmwylab2}.
	\end{equation}
	The transformation resonance frequency is $D \pm g{\mu _B}B\cos \theta  \pm {\omega _r}\left( {1 - \cos \theta } \right)$ between the ${\left| {{m_s} = 0} \right\rangle _{lab}}$ state and ${\left| {{m_s} =  \pm 1} \right\rangle _{lab}}$ state. The corresponding frequency shift duo to the Berry phase also is  $\pm {\omega _r} \cos \theta $, and the frequency shift induced by the rotational Doppler effect is $\omega _r$. Similar to the case of zero external magnetic field, the predominant transition probability arises from the longitudinal component, characterized by a frequency shift of $\mp {\omega _r}\cos \theta$ due to the Berry phase.}

\rw{Supplementary Fig. \ref{fig:s6}(c) shows the frequency shift induced by the Berry phase in a levitated nanodiamond rotating counterclockwisely (viewed from the positive $z$ direction unless otherwise specified). The external magnetic field along the z direction is about 100 G. The resonance frequency transition between ${\left| {{m_s} = 0} \right\rangle _{lab}}$ state and ${\left| {{m_s} =  + 1} \right\rangle _{lab}}$ state decreases with an increasing of the rotation frequency, in contrast to the behavior observed in the levitated nanodiamond rotating clockwise. The red curve is the theoretical calculation for the angle of $\theta = 21.5^\circ$ between the NV axis and the rotating axis. The experimental data is in excellent agreement with the theoretical calculation, suggesting a consistent orientation of the NV centers at various rotation frequencies.}

\subsection{Pseudo-magnetic field due to rotation}
\rw{The Berry phase observed in the laboratory frame is equivalent to the pseudo-magnetic field (called the Barnett field in \cite{Chudo2021}) in the rotational frame. In our experiment, the mircowave source is fixed in the laboratory frame. Only the levitated nanodiamond is rotating. Thus, we observe the effect of the Berry phase \cite{Chudo2021}. It will be beneficial to also consider this system in the rotational frame. The electron spin resonance frequency shift of the rotating NV center involves the combination of the pseudo-magnetic field  and the rotational Doppler effect in the rotating frame. As expressed in Eq.~\ref{eq:hrot}, the Hamiltonian of the pseudo-magnetic field in the rotating frame, induced by the rotation of a diamond particle, can be given by
	\begin{equation}
		{H_{{\omega _r}}} = \hbar \left( {\begin{array}{*{20}{c}}
				{ - {\omega _r}\cos \theta }&{\frac{{{\omega _r}\sin \theta }}{{\sqrt 2 }}}&0\\
				{\frac{{{\omega _r}\sin \theta }}{{\sqrt 2 }}}&0&{\frac{{{\omega _r}\sin \theta }}{{\sqrt 2 }}}\\
				0&{\frac{{{\omega _r}\sin \theta }}{{\sqrt 2 }}}&{{\omega _r}\cos \theta }
		\end{array}} \right) \approx \hbar \left( {\begin{array}{*{20}{c}}
				{ - {\omega _r}\cos \theta }&0&0\\
				0&0&0\\
				0&0&{{\omega _r}\cos \theta }
		\end{array}} \right)
		\label{eq:hwr},
	\end{equation}
	where the off-diagonal terms also can be ignored. So the Hamiltonian of the NV center in the rotating frame can be expressed as
	\begin{equation}
		{H_{rot}} = \hbar \left( {\begin{array}{*{20}{c}}
				{D + g{\mu _B}B\cos \theta  - {\omega _r}\cos \theta }&0&0\\
				0&0&0\\
				0&0&{D - g{\mu _B}B\cos \theta  + {\omega _r}\cos \theta }
		\end{array}} \right)
		\label{eq:hrot2}.
	\end{equation}
	The corresponding eigenvalues are $\hbar \left( {D + g{\mu _B}B\cos \theta  - {\omega _r}\cos \theta } \right)$, 0, $\hbar \left( {D - g{\mu _B}B\cos \theta  + {\omega _r}\cos \theta } \right)$ for spin state ${\left| + 1 \right\rangle _{rot}}$, ${\left| 0 \right\rangle _{rot}}$ and ${\left| - 1 \right\rangle _{rot}}$, respectively.}

\rw{The Hamiltonian describing the interaction of the microwave term with the NV center in the rotating frame can be expressed as
	\begin{equation}
		\begin{array}{l}
			{H_{MW,rot}} = U{H_{MW,lab}}{U^\dag } = {e^{i\theta {S_y}}}{e^{i\phi {S_z}}}{H_{MW,lab}}{e^{ - i\phi {S_z}}}{e^{ - i\theta {S_y}}}\\
			= g{\mu _B}{B_{MW}}\cos \left( {{\omega _{MW}}t} \right){e^{i\theta {S_y}}}{e^{i\phi {S_z}}}\left( {{S_z}\cos \theta ' + {S_y}\sin \theta '} \right){e^{ - i\phi {S_z}}}{e^{ - i\theta {S_y}}} = {H_{MW,z,rot}} + {H_{MW,y,rot}}
		\end{array}
		\label{eq:hmwrot},
	\end{equation}
	where the longitudinal component is ${H_{MW,z,rot}} = g{\mu _B}{B_{MW}}\cos \left( {{\omega _{MW}}t} \right){e^{i\theta {S_y}}}{e^{i\phi {S_z}}}{S_z}\cos \theta '{e^{ - i\phi {S_z}}}{e^{ - i\theta {S_y}}}$, and the transverse component is ${H_{MW,y,rot}} = g{\mu _B}{B_{MW}}\cos \left( {{\omega _{MW}}t} \right){e^{i\theta {S_y}}}{e^{i\phi {S_z}}}{S_y}\sin \theta '{e^{ - i\phi {S_z}}}{e^{ - i\theta {S_y}}}$. The expected value of the spin state, interacting with the longitudinal component of microwave, can be expressed as
	\begin{equation}
		\begin{array}{l}
			_{rot}\left\langle { \pm 1} \right|{e^{i{H_{rot}}t/\hbar }}{H_{MW,z,rot}}{e^{ - i{H_{rot}}t/\hbar }}{\left| 0 \right\rangle _{rot}}\\
			= g{\mu _B}{B_{MW}}\cos \theta '\cos {\left( {{\omega _{MW}}t} \right)_{rot}}\left\langle { \pm 1} \right|{e^{i{H_{rot}}t/\hbar }}{e^{i\theta {S_y}}}{e^{i\phi {S_z}}}{S_z}{e^{ - i\phi {S_z}}}{e^{ - i\theta {S_y}}}{e^{ - i{H_{rot}}t/\hbar }}{\left| 0 \right\rangle _{rot}}\\
			= g{\mu _B}{B_{MW}}\cos \theta '\cos \left( {{\omega _{MW}}t} \right){e^{i\left( {{E_{ \pm 1,rot}} - {E_{0,rot}}} \right)t/\hbar }}_{rot}\left\langle { \pm 1} \right|{e^{i\theta {S_y}}}{e^{i\phi {S_z}}}{S_z}{e^{ - i\phi {S_z}}}{e^{ - i\theta {S_y}}}{\left| 0 \right\rangle _{rot}}\\
			= \frac{1}{2}g{\mu _B}{B_{MW}}\cos \theta '\left( {{e^{i{\omega _{MW}}t}} + {e^{ - i{\omega _{MW}}t}}} \right){e^{i\left( {D \pm g{\mu _B}B\cos \theta  \mp {\omega _r}\cos \theta } \right)t}}_{rot}\left\langle { \pm 1} \right|{e^{i\theta {S_y}}}{S_z}{e^{ - i\theta {S_y}}}{\left| 0 \right\rangle _{rot}}\\
			= \frac{1}{2}g{\mu _B}{B_{MW}}\cos \theta '\left[ {{e^{i\left( {{\omega _{MW}} + D \pm g{\mu _B}B\cos \theta  \mp {\omega _r}\cos \theta } \right)t}} + {e^{i\left( { - {\omega _{MW}}D \pm g{\mu _B}B\cos \theta  \mp {\omega _r}\cos \theta } \right)t}}} \right]\\
			{ \times _{rot}}\left\langle { \pm 1} \right|{e^{i\theta {S_y}}}{S_z}{e^{ - i\theta {S_y}}}{\left| 0 \right\rangle _{rot}}\\
			\approx \frac{1}{2}g{\mu _B}{B_{MW}}\cos \theta '{e^{i\left( { - {\omega _{MW}}D \pm g{\mu _B}B\cos \theta  \mp {\omega _r}\cos \theta } \right)t}}_{rot}\left\langle { \pm 1} \right|{e^{i\theta {S_y}}}{S_z}{e^{ - i\theta {S_y}}}{\left| 0 \right\rangle _{rot}}
		\end{array}
		\label{eq:hmwrz},
	\end{equation}
	where the $E_{m_s,rot}$ is the eigenvalue of the Hamiltonian $H_{rot}$ for the spin state ${\left| m_s \right\rangle _{rot}}$ in the rotating frame. The transformation resonance frequency is $D \pm g{\mu _B}B\cos \theta  \mp {\omega _r}\cos \theta $, and the last term $\mp {\omega _r}\cos \theta$ is the frequency shift due to the pseudo-magnetic field. The transverse component can be expressed as
	\begin{equation}
		\begin{array}{l}
			_{rot}\left\langle { \pm 1} \right|{e^{i{H_{rot}}t/\hbar }}{H_{MW,y,rot}}{e^{ - i{H_{rot}}t/\hbar }}{\left| 0 \right\rangle _{rot}}\\
			= g{\mu _B}{B_{MW}}\sin \theta '\cos {\left( {{\omega _{MW}}t} \right)_{rot}}\left\langle { \pm 1} \right|{e^{i{H_{rot}}t/\hbar }}{e^{i\theta {S_y}}}{e^{i\phi {S_z}}}{S_y}{e^{ - i\phi {S_z}}}{e^{ - i\theta {S_y}}}{e^{ - i{H_{rot}}t/\hbar }}{\left| 0 \right\rangle _{rot}}\\
			= g{\mu _B}{B_{MW}}\sin \theta '\cos \left( {{\omega _{MW}}t} \right){e^{i\left( {{E_{ \pm 1,rot}} - {E_{0,rot}}} \right)t/\hbar }}_{rot}\left\langle { \pm 1} \right|{e^{i\theta {S_y}}}{e^{i\phi {S_z}}}{S_y}{e^{ - i\phi {S_z}}}{e^{ - i\theta {S_y}}}{\left| 0 \right\rangle _{rot}}\\
			= \frac{1}{2}g{\mu _B}{B_{MW}}\sin \theta '\left( {{e^{i{\omega _{MW}}t}} + {e^{ - i{\omega _{MW}}t}}} \right){e^{i\left( {D \pm g{\mu _B}B\cos \theta  \mp {\omega _r}\cos \theta } \right)t}}\\
			{ \times _{rot}}\left\langle { \pm 1} \right|{e^{i\theta {S_y}}}\frac{1}{{2i}}\left( {{S_ + }{e^{i{\omega _r}t}} - {S_ - }{e^{ - i{\omega _r}t}}} \right){e^{ - i\theta {S_y}}}{\left| 0 \right\rangle _{rot}}
		\end{array}
		\label{eq:hmwyrot}.
	\end{equation} 
	The expected values are written as
	\begin{equation}
		\begin{array}{l}
			_{rot}\left\langle { + 1} \right|{e^{i{H_{rot}}t/\hbar }}{H_{MW,y,rot}}{e^{ - i{H_{rot}}t/\hbar }}{\left| 0 \right\rangle _{rot}}\\
			= \frac{1}{{4i}}g{\mu _B}{B_{MW}}\sin \theta '\left[ {{e^{i\left( {{\omega _{MW}} + D + g{\mu _B}B\cos \theta  - {\omega _r}\cos \theta  + {\omega _r}} \right)t}} + {e^{i\left( { - {\omega _{MW}} + D + g{\mu _B}B\cos \theta  - {\omega _r}\cos \theta  + {\omega _r}} \right)t}}} \right]\\
			{ \times _{rot}}\left\langle { + 1} \right|{e^{i\theta {S_y}}}{S_ + }{e^{ - i\theta {S_y}}}{\left| 0 \right\rangle _{rot}}\\
			\approx \frac{1}{{4i}}g{\mu _B}{B_{MW}}\sin \theta '{e^{i\left( { - {\omega _{MW}} + D + g{\mu _B}B\cos \theta  - {\omega _r}\cos \theta  + {\omega _r}} \right)t}}_{rot}\left\langle { + 1} \right|{e^{i\theta {S_y}}}{S_ + }{e^{ - i\theta {S_y}}}{\left| 0 \right\rangle _{rot}}
		\end{array}
		\label{eq:hmwyrot1},
	\end{equation}
	\begin{equation}
		\begin{array}{l}
			_{rot}\left\langle { - 1} \right|{e^{i{H_{rot}}t/\hbar }}{H_{MW,y,rot}}{e^{ - i{H_{rot}}t/\hbar }}{\left| 0 \right\rangle _{rot}}\\
			=  - \frac{1}{{4i}}g{\mu _B}{B_{MW}}\sin \theta '\left[ {{e^{i\left( {{\omega _{MW}} + D - g{\mu _B}B\cos \theta  + {\omega _r}\cos \theta  - {\omega _r}} \right)t}} + {e^{i\left( { - {\omega _{MW}} + D - g{\mu _B}B\cos \theta  + {\omega _r}\cos \theta  - {\omega _r}} \right)t}}} \right]\\
			{ \times _{rot}}\left\langle { - 1} \right|{e^{i\theta {S_y}}}{S_ - }{e^{ - i\theta {S_y}}}{\left| 0 \right\rangle _{rot}}\\
			\approx  - \frac{1}{{4i}}g{\mu _B}{B_{MW}}\sin \theta '{e^{i\left( { - {\omega _{MW}} + D - g{\mu _B}B\cos \theta  + {\omega _r}\cos \theta  - {\omega _r}} \right)t}}_{rot}\left\langle { + 1} \right|{e^{i\theta {S_y}}}{S_ + }{e^{ - i\theta {S_y}}}{\left| 0 \right\rangle _{rot}}
		\end{array}
		\label{eq:hmwyrot2}.
	\end{equation}
	The transformation resonance frequency  between the ${\left| {{m_s} = 0} \right\rangle _{rot}}$ state and ${\left| {{m_s} =  \pm 1} \right\rangle _{rot}}$ state is $D \pm g{\mu _B}B\cos \theta  \pm {\omega _r}\left( {1 - \cos \theta } \right)$. The frequency shift of $\pm {\omega _r}\left( {1 - \cos \theta } \right)$ is induced by both the pseudo-magnetic field and the rotational Doppler effect \cite{Chudo2021} in the rotating frame.}

\section{Quantum measurement of levitated diamond NV centers}

\begin{figure*}
	\includegraphics[width=\textwidth]{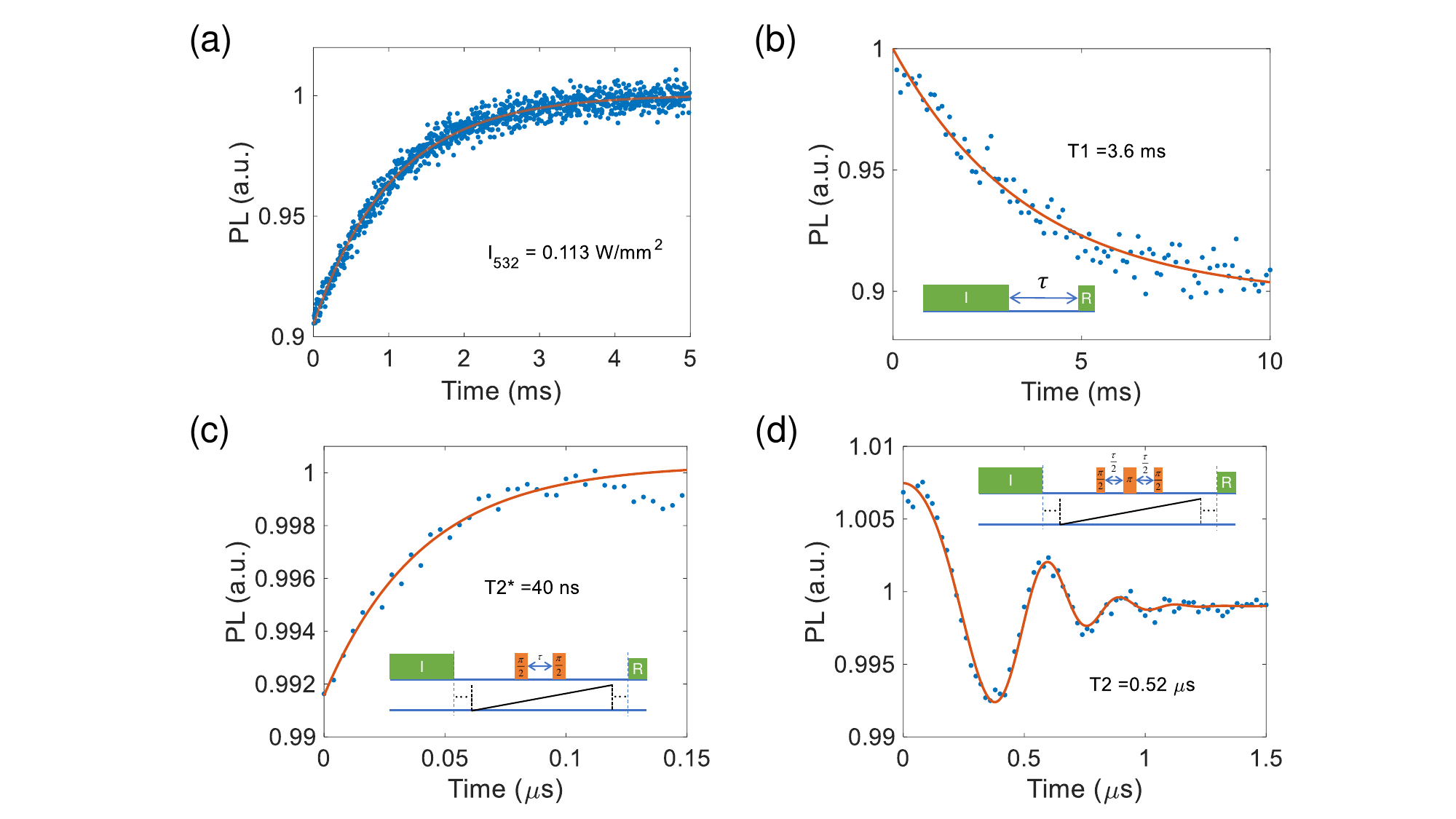}
	\caption{\label{fig:s7} Quantum measurement of NV centers in a levitated nanodiamond. (a) Initialization time of levitated Nv centers when the 532 nm laser intensity is 0.113 W/mm$^2$. The initialization time is 1.05 ms. (b) Experimental result of $T_1$ measurement of the levitated nanodiamond. \rw{The insert is the sequence of $T_1$ measurement.} The $T_1$ is 3.6 ms. \rw{(c) Ramsey measurement and its corresponding sequence with a $T_2^*$ of 40 ns. (d) Spin Echo measurement and its corresponding sequence with a $T_2$ of 0.52 $\mu$s. The oscillation is induced by the misalignment of the magnetic field with the rotation axis. In the sequences, the green and orange regions represent the pulses of the 532 nm laser and the microwave, respectively. The black lines are the rotation phase.}}
\end{figure*}

The power of 532 nm laser for NV initialization is very weak to minimize laser heating in high vacuum, leading to a long NV polarization time. The measured initialization time, shown in Supplementary Fig. \ref{fig:s7}(a), is 1.05 ms at the 532 nm laser intensity of 0.113 W/mm$^2$. We measure the spin relaxation time ($T_1$) of the levitated nanodiamond in Supplementary Fig. \ref{fig:s7}(c), indicating $T_1 = 3.60$ ms. It is three times longer than the initialization time. \rw{The $T_2^*$ and $T_2$ are measured, as shown in  Supplementary Fig. \ref{fig:s7}(c) and (d). The $T_2^*$ is 40~ns and the $T_2$ is 0.52~$\mu$s. In the spin echo measurement, the oscillation is a result of the misalignment of the magnetic field with the rotation axis \cite{Maze2008}. }

Due to the $\Omega$-shape design of the microwave antenna, the orientation of the magnetic component of microwave is located in $yz$-plane and slightly away from the $z$ axis with an angle about $\theta ' = 8.5 ^\circ$. Thus, the effective magnetic field $B_{MW}^ \bot$ of the microwave acting on NV spins  keeps varying at different rotation phase $\phi (t)$ of the levitated nanodiamond. We set \rw{the direction of} a NV spin to be \rw{ ${\bf{n}}_{NV} = \left( { \sin \theta, 0,\cos \theta } \right)$} in the \rw{$xz$}-plane at initial time ($t = 0$), and then rotate it \rw{around} the $z$ axis. The rotation matrix is:
\begin{eqnarray}
	{{\bf{r}}_z} = \left( {\begin{array}{*{20}{c}}
			{\cos \phi \left( t \right)}&{ - \sin \phi \left( t \right)}&0\\
			{\sin \phi \left( t \right)}&{\cos \phi \left( t \right)}&0\\
			0&0&1
	\end{array}} \right)
	\label{eq:s14}.
\end{eqnarray}
After a rotation time of $t$, the \rw{direction of the} NV spin is changed to
\begin{eqnarray}
	\rw{{\bf{n}'}_{NV} = \left( {\cos \phi \left( t \right)\sin \theta ,\sin \phi \left( t \right)\sin \theta ,\cos \theta } \right)}
	\label{eq:s15}.
\end{eqnarray}
The unit vector of the direction of the magnetic field of microwave is ${\bf{n}}_{MW} = \left( {0, -\sin \theta ' ,\cos \theta '} \right)$. So the angle between the NV spin and the microwave is $\arccos \left( {\cos \theta \cos \theta ' - \rw{\sin}\phi \left( t \right)\sin \theta \sin \theta '} \right)$. The effective magnetic field of microwave is
\begin{eqnarray}
	B_{MW}^ \bot  = {B_{MW}}\sqrt {1 - {{\left( {\cos \theta \cos \theta ' - \rw{\sin}\phi \left( t \right)\sin \theta \sin \theta '} \right)}^2}} 
	\label{eq:s16}.
\end{eqnarray}
Rabi frequency is proportional to the effective magnetic field of microwave, ${\Omega _{Rabi}} \propto B_{MW}^ \bot$. Therefore, it is necessary to synchronize the measurement cycle and the rotation signal of the levitated nanodiamond, and apply microwave pulse at the same rotation phase in repeated Rabi oscillation measurements.

\begin{figure*}
	\includegraphics[width=\textwidth]{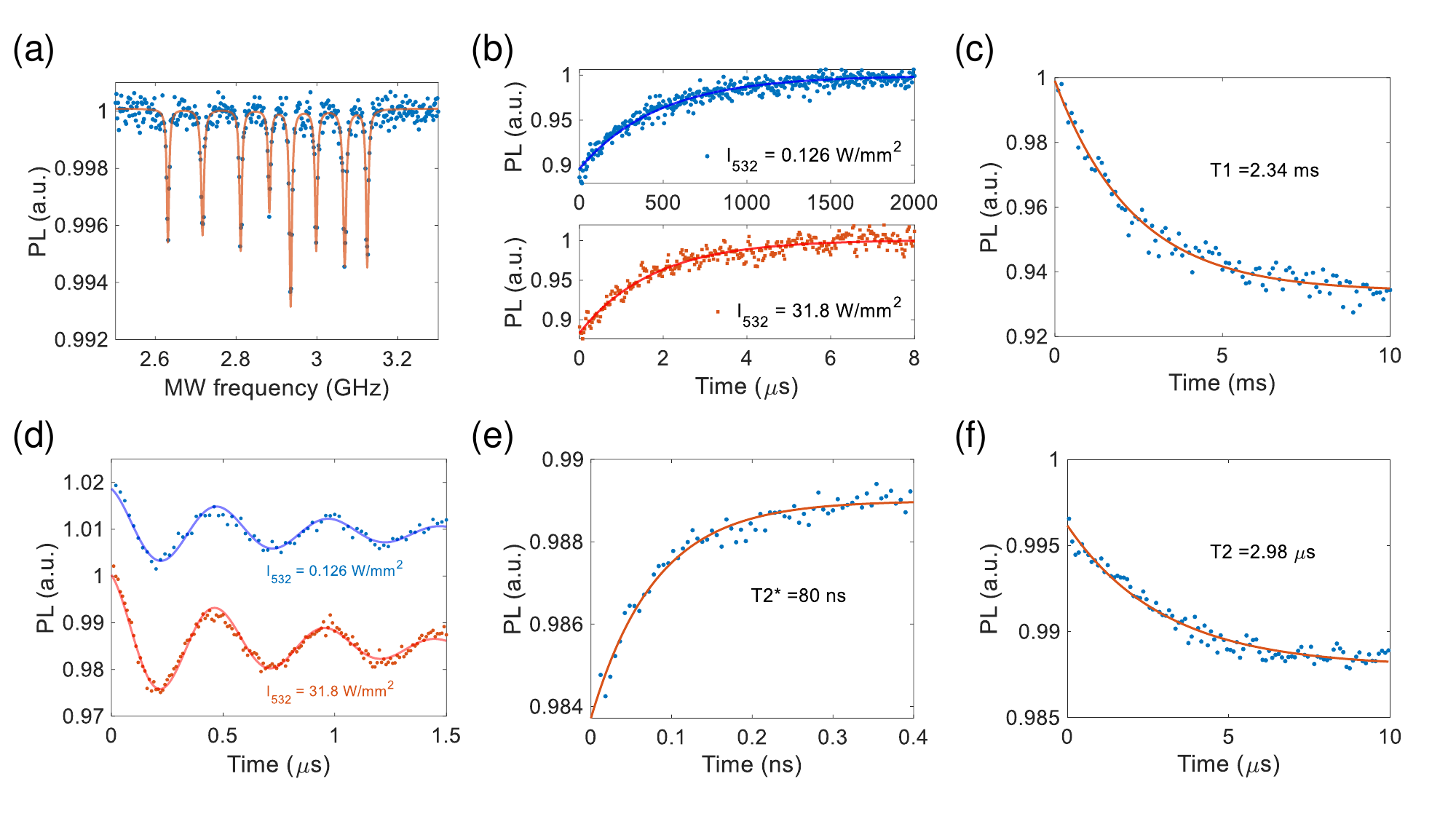}
	\caption{\label{fig:s8} Quantum measurement of NV centers in a nanodiamond fixed on a glass. (a) ODMR of the fixed NV centers. (b) Initialization of the nanodiamond Nv centers when the 532 nm laser intensity is 0.126 W/mm$^2$ (blue circles) or 31.8 W/mm$^2$ (red squares). The initialization times are 0.469 ms and 1.77 $\mu$s, respectively. (c) Experimental $T_1$ measurement of the nanodiamond. The $T_1$ is 2.34 ms. (d) Rabi oscillations of the nanodiamond when the 532 nm laser intensity is 0.126 W/mm$^2$ (blue circles) or 31.8 W/mm$^2$ (red squares). The Rabi frequencies are both 1.99 MHz and the decay times $T_2^{\mathrm{rabi}}$ are 0.845 $\mu$s and 0.904 $\mu$s, respectively. \rw{(e) Experimental Ramsey measurement. The $T_2^*$ of the nanodiamond NV centers fixed on a glass is 80 ns. (f) Spin Echo measurement, which indicates a $T_2$ of 2.98 $\mu$s.}}
\end{figure*}

\section{Nanodiamond fixed on glass}

To compare with a levitated nanodiamond, we carry out the quantum measurement of a nanodiamond fixed on a glass cover slip, with a thickness of 300 $\mu$m. The glass cover slip is placed at the center of the surface ion trap to keep the direction and power of microwave unchanged compared to a levitated nanodiamond. Supplementary Fig. \ref{fig:s8}(a) is the ODMR of the nanodiamond. The linewidth is smaller than that of a levitated nanodiamond. The initialization time are measured at $I_{532} = 0.126$ W/mm$^2$ (blue circles) and $I_{532} = 31.8$ W/mm$^2$ (red squares in Supplementary Fig. \ref{fig:s8}(b)) for comparison. The initialization times are 0.469 ms and 1.77 $\mu$s, respectively. T1 of the nanodiamond on glass surface is 2.34 ms (Supplementary Fig. \ref{fig:s8}(c)), which is close to that of a levitated one. We measure Rabi oscillation at weak 532 nm laser ($I_{532} = 0.126$ W/mm$^2$), which is similar to the intensity used for a levitated nanodiamond in high vacuum (Supplementary Fig. \ref{fig:s8}(d)). We get the similar result with high intensity of 532 nm laser ($I_{532} = 31.8$ W/mm$^2$). The Rabi frequency is 1.99 MHz and the decay time $T_2^{\mathrm{Rabi}}$ are 0.845 $\mu$s and 0.904 $\mu$s, respectively. \rw{Supplementary Fig.~\ref{fig:s8}(e) shows the Ramsey measurement, while Supplementary Fig.~\ref{fig:s8}(f) is the spin echo measurement. The corresponding values for $T_2^*$ and $T_2$ are 80 ns and 2.98 $\mu$s, respectively. }

\end{document}